\documentclass[aps,prb,twocolumn,10pt,longbibliography]{revtex4-1}
\usepackage{amsmath}
\usepackage{amsfonts}
\usepackage{amssymb}
\usepackage{mathtools}
\usepackage{bm}
\usepackage{cases}
\usepackage[version=4]{mhchem}
\usepackage{graphicx}
\usepackage[caption=false]{subfig}
\allowdisplaybreaks%

\begin{document}
\title{Pr$_2$Ir$_2$O$_7$: when Luttinger semimetal meets Melko-Hertog-Gingras spin ice state}
\author{Xu-Ping Yao}
\affiliation{Department of Physics, 
Center for Field Theory and Particle Physics, 
State Key Laboratory of Surface Physics,
Fudan University, Shanghai 200433, China}
\author{Gang Chen}
\email{gangchen.physics@gmail.com}
\affiliation{Department of Physics, 
Center for Field Theory and Particle Physics, 
State Key Laboratory of Surface Physics,
Fudan University, Shanghai 200433, China}
\affiliation{Collaborative Innovation Center of Advanced 
Microstructures, Nanjing, 210093, China}

\date{\today}
    
\begin{abstract}
We study the band structure topology and engineering from the interplay
between local moments and itinerant electrons in the context 
of pyrochlore iridates. For the metallic iridate Pr$_2$Ir$_2$O$_7$,
the Ir $5d$ conduction electrons interact with the Pr $4f$ local moments 
via the $f$-$d$ exchange. While the Ir electrons form a 
Luttinger semimetal, the Pr moments can be tuned into an ordered 
spin ice with a finite ordering wavevector, dubbed ``Melko-Hertog-Gingras''  
state, by varying Ir and O contents. We point out that the ordered 
spin ice of the Pr local moments generates an internal magnetic field 
that reconstructs the band structure of the Luttinger semimetal. 
Besides the broad existence of Weyl nodes, we predict that the 
magnetic translation of the ``Melko-Hertog-Gingras'' state for 
the Pr moments protects the Dirac band touching at certain time 
reversal invariant momenta for the Ir conduction electrons.  
We propose the magnetic fields to control the Pr magnetic structure 
and thereby indirectly influence the topological and other properties 
of the Ir electrons. Our prediction may be immediately tested in the 
ordered Pr$_2$Ir$_2$O$_7$ samples. We expect our work to stimulate a 
detailed examination of the band structure, magneto-transport, and 
other properties of Pr$_2$Ir$_2$O$_7$.
\end{abstract}

\maketitle

\section{Introduction}
\label{sec1}
    
The study of the electron band structure topology has attracted 
a significant attention since the proposal and discovery of 
topological insulators~\cite{KaneReview,ShouchengReview,VishwanathReview,FuTCI}. 
The fundamental aspect between the 
topological protection and the band structure has been well
understood. More practically, there is a growing 
effort that proposes experimental schemes such as the strain
or magnetic dopings to control or engineer the band structure 
topology. Along this line, a great success was achieved in 
the discovery of quantum anomalous Hall effect in the 
magnetically doped Bi$_2$Se$_3$ materials~\cite{Yu61,Chang167,QAHE}. 
Quantum materials, that contain extra degrees of freedom besides
the nearly free electrons, would be ideal for the practical 
purpose to control the band structure properties. In this paper, 
we illustrate this general idea and specifically study 
the Pr-based pyrochlore iridate. We show that the $4f$ 
local moment of the Pr ions and their impacts on the Ir conduction 
electrons provide a natural setup to explore the band structure 
engineering via the coupling between these two degrees of freedom. 

Pyrochlore iridates, \ce{R2Ir2O7}, have received a considerable attention 
in recent years partly because the $5d$ electrons of the Ir pyrochlore 
system provide an interesting arena to explore the correlation effects 
in the strong spin-orbit-coupled matters~\cite{PesinBalents,JPSJ.70.2880,Pyrochloreiridate2002}. 
Aligned with this original motivation, many interesting phases and phenomena, 
including topological Mott insulator, axion insulator, Weyl semimetal, 
Luttinger-Abrikosov-Beneslavskii non-Fermi liquid, {\sl{et al}}, 
have been proposed~\cite{WanXG,BJYang,EGMoon,YBKim1,YBKim2,
ChenHermele,LeeArun,SavaryMoon,WCKB,Millis,MoessnerSchnyder}. 
Despite the fruitful achievements since the original motivation~\cite{PhysRevB.85.205104,
PhysRevB.85.245109,PhysRevB.83.180402,PhysRevB.86.014428,PhysRevB.85.174441,
PhysRevB.90.054419,PhysRevB.90.235110,PhysRevB.93.104422}, 
the role of the rare-earth local moments in this system has not been 
extensively studied except few works~\cite{ChenHermele,LeeArun,PhysRevB.87.125147}. 
Recent experiments~\cite{Nd2Ir2O7,Machida} in \ce{Nd2Ir2O7} and \ce{Pr2Ir2O7} 
do suggest the importance of the local moments and the coupling between 
the local moments and the Ir conduction electrons. 

Our work is mainly inspired by the experiments on \ce{Pr2Ir2O7}. 
Depending on the stoichiometry, the \ce{Pr2Ir2O7} samples show 
rather distinct behaviors. While the early samples remain metallic 
and paramagnetic down to the lowest temperature~\cite{PhysRevLett.96.087204,Gegenwart,Machida}, 
recent samples with different iridium and oxygen contents develop 
an antiferromagnetic Ising order in the Pr subsystem with a $2\pi(001)$ 
ordering wavevector~\cite{PhysRevB.92.054432}. 
This particular order is a state within the spin ice manifold and coincides 
with the classical spin ground state of classical dipolar spin model on the 
pyrochlore lattice that was found by Melko, Hertog, and Gingras in 
Ref.~\onlinecite{MelkoHertogGingras}. Although the physical origin 
of this order in \ce{Pr2Ir2O7} differs from the classical and dipolar 
interaction in Ref.~\onlinecite{MelkoHertogGingras}, we refer this 
Ising order as ``Melko-Hertog-Gingras'' state or order. Since the Pr 
local moment was argued to fluctuate within the ``2-in 2-out'' spin ice 
manifold in the paramagnetic samples and the ``Melko-Hertog-Gingras'' state 
of the ordered samples is a particular antiferromagnetic state within the 
spin ice manifold, it was proposed by one of us that, the Pr subsystem is 
proximate to a quantum phase transition from the $U(1)$ quantum spin liquid 
to the ``Melko-Hertog-Gingras'' order via a confinement transition by 
proliferating the ``magnetic monopoles''~\cite{GangChen2016}.

Besides the interesting aspects of the Pr local moments, the Ir conduction 
electron was shown to display interesting phenomena. Recent works have 
identified the presence of a quadratic band touching at the $\Gamma$ point 
for the Ir $5d$ electrons~\cite{Kondo2015,Onoda2015,Armitage2017}. Theoretical 
works have considered the long-range Coulomb interaction for the Luttinger 
semimetal phase of the Ir subsystem~\cite{EGMoon}. 
These efforts surely fall into the original motivation of searching for 
correlation physics in strong spin-orbit-coupled matter~\cite{PesinBalents,WCKB} 
and provide an important understanding of the rich physics in this material. 
The purpose of this work is to deviate from the intense efforts on the correlation 
physics of the Ir subsystems, and is instead to understand the interplay 
between the Ir conduction electrons and the Pr local moments. As is already 
pointed out in Ref.~\onlinecite{GangChen2016}, the large and finite ordering wavevector 
of the Pr local moments and the quadratic band touching of the Ir $5d$ electrons 
suppress the Yukawa coupling between the Pr magnetic order and the Ir particle-hole  
excitation near the (small) Fermi surface or $\Gamma$ point. Therefore, the Ir 
electron near the Fermi surface does not modify the critical and long-distance 
properties of the Pr local moments at the lowest order, though it is thought 
that the phase transition of the Pr local moment was induced by the modified 
Ruderman-Kittel-Kasuya-Yosida (RKKY) interaction mediated by the Ir electrons~\cite{GangChen2016}. 
The opposite, however, is not true. The Luttinger semimetal with a quadratic 
band touching is a parent state of various topological phases such as Weyl 
semimetal and topological insulator~\cite{Kondo2015,Bernevig1757,PhysRevLett.106.126803}. 
The coupling to the Pr local moment naturally provides 
such a perturbation to the parabolic semimetal. In this work, 
we focus on the Ising ordered phase and explain the 
effect of the Pr magnetism on the Ir conduction electrons.

In \ce{Eu2Ir2O7} and other pyrochlore iridates~\cite{Nd2Ir2O7,PhysRevB.83.180402,PhysRevB.85.245109}, the Ir subsystem experiences 
a metal-insulator transition by developing an all-in all-out magnetic order 
with the ordering wavevector ${{\boldsymbol Q}={\boldsymbol 0}}$, and it is believed 
that the magnetic order is driven by the correlation of the Ir $5d$ electrons. 
For \ce{Pr2Ir2O7}, the magnetic unit cell of \ce{Pr2Ir2O7} is twice the 
size of the crystal unit cell, and it is the Pr local moment that develops 
the magnetic order. The exchange field, that is experienced by the Ir 
conduction electron and generated by the ``Melko-Hertog-Gingras'' order
of Pr moments, is thus very different from other pyrochlore iridates. 
Therefore, we construct a minimal model to incorporate the coupling 
and interactions of the relevant microscopic degrees of freedom. 
This model, as we introduce in Sec.~\ref{sec2}, naturally captures 
the physics that we describe above. We find that, the exchange field 
enlarges the unit cell of the Ir subsystems, and couples the electrons/holes 
near the $\Gamma$ point with the electrons/holes near the ordering 
wavevector ${{\boldsymbol Q}=2\pi(001)}$. The combination of the 
time reversal operation and the elementary lattice translation by
$(1/2,0,1/2)$ or $(0,1/2,1/2)$ remains to be an (anti-unitary) 
symmetry of the ``Melko-Hertog-Gingras'' state. Using this symmetry, 
we demonstrate that there exist Dirac band touchings at the 
high symmetry momenta. Our explicit calculation with the realistic 
model confirms these band touchings. In addition, we find the 
existence of Weyl nodes in the Ir band structure 
due to the breaking of the time reversal by the Pr magnetic order. 
Unlike the symmetry protected Dirac band touchings, the Weyl nodes 
are not symmetry protected and are instead topologically stable. 

Apart from the immediate effect on the Ir conduction electron from the Pr 
Ising magnetic order, we further explore the role of the external magnetic field. 
It is noticed that, the external magnetic field primarily couples to the Pr 
local moments rather than to the conduction electron. This is because every 
Pr local moment couples to the external magnetic field, while for the Ir 
conduction electrons only small amount of electrons on the Fermi surface 
couple to the magnetic field, not to say, there is a vanishing density 
of states right at the $\Gamma$ point of the quadratic band touching. 
The Zeeman coupling to the Pr local moment would simply favor a   
${{\boldsymbol Q}={\boldsymbol 0}}$ state and thus competes with the exchange 
interaction of the Pr subsystem. The combination of the magnetic field 
and the Pr exchange coupling generates several different magnetic states 
for the Pr local moments. These magnetic orders create distinct exchange 
fields on the Ir conduction electrons and thereby gives new reconstructions 
of the conduction electron band structure. From the symmetry point of view, 
the Dirac band touchings at the time reversal invariant momenta are no longer 
present in the magnetic field. We further find that the Weyl nodes
exist broadly when the magnetic field is applied to the system. 
This provides a feasible experimental scheme to engineer the band 
structure properties of the Ir itinerant electrons.  

The following part of the paper is organized as follows. In Sec.~\ref{sec2}, 
we introduce the microscopic Hamiltonian for the Ir subsystem and the $f$-$d$ 
exchange between the Ir subsystem and the Pr subsystem. In Sec.~\ref{sec3}, 
we include the antiferromagnetic Ising order of the Pr local moments and 
study the reconstruction of the Ir band structure under this magnetic order.  
In Sec.~\ref{sec4}, we further explore the interplay between the Zeeman 
coupling, the Pr exchange coupling and the Ir band structure, and point 
out that the external field can be used to engineer the topological band 
structure. Finally in Sec.~\ref{sec5}, we conclude with a discussion and 
propose various experiments to confirm our prediction.

\section{Microscopic model}
\label{sec2} 
 
We here propose the minimal microscopic model for Pr$_2$Ir$_2$O$_7$ and 
explain the limitation of the model. The approximation in the minimal model
is further justified and designed to reveal the physics that we want to discuss. 
The full Hamiltonian of this system should contain the following 
ingredients~\cite{ChenHermele},
\begin{eqnarray}
H = H_{tb} + H_{ex} + H_{fd} + H_{\text{Zeeman}}, 
\label{eq1}
\end{eqnarray}
where $H_{tb}$ is the tight-binding model of the Ir conduction electron,
$H_{ex}$ is the interaction between the Pr local moments and originates
from the superexchange process and the dipolar interaction, $H_{fd}$ is 
the coupling between the Pr local moment with the spin density of the 
Ir conduction electrons, and the $H_{\text{Zeeman}}$ defines the Zeeman 
coupling of the Pr local moment to the external magnetic field. 

\subsection{Ir subsystem}

We start with the tight-binding model for the Ir conduction electrons. 
The Ir$^{4+}$ ion has a $5d^5$ electron configuration, and these five 
electrons occupy the $t_{2g}$ orbitals. The atomic spin-orbit coupling 
splits the six-fold degenerate spin and orbital states in the $t_{2g}$ 
manifold into the lower ${j=3/2}$ quadruplets and the upper ${j=1/2}$ 
doublets. Due to the lattice geometry of the pyrochlore system, 
the $t_{2g}$ orbitals and the effective spin ${\boldsymbol J}$  
are defined in the local coordinate system of the IrO$_6$ octahedron.
For the Ir$^{4+}$ ion, the lower ${j=3/2}$ quadruplets are fully filled, 
and the upper ${j=1/2}$ doublets are halfly filled~\cite{BJKim,PhysRevB.78.094403,Khaliullin,PesinBalents}. It was shown that, the pyrochlore 
iridate band structure near the Fermi level is well approximated by a 
tight-binding model based on the ${j=1/2}$ doublets~\cite{PesinBalents,BJYang,YBKim1}. 
The model is given as
\begin{equation}
H_{tb} = \sum_{i,j \in \text{Ir}}\sum_{\alpha\beta}  
t^{}_{ij,\alpha\beta} d^{\dagger}_{i\alpha} d^{}_{j\beta},
\end{equation}
where $d^{\dagger}_{i\alpha}$ ($d^{}_{i\alpha}$) creates    
(annihilates) an electron with an effective spin $\alpha$ 
in the ${j=1/2}$ doublet. The hopping $t^{}_{ij,\alpha\beta}$
includes both the direct electron hoppings ($t_{\sigma}$ and $t_{\pi}$) 
between the nearest-neighbor Ir ions and the indirect electron hopping 
($t_{id}$) through the intermediate oxygen. It has been shown~\cite{YBKim1} 
that in the regime ${-1.67t_{id} < t_{\sigma} < -0.67t_{id}}$
and ${t_{\pi} = -2t_{\sigma}}/3$, 
the system becomes a Luttinger semimetal with a quadratic band 
touching at the $\Gamma$ point. This quadratic band touching 
is protected by the cubic lattice symmetry~\cite{BJYang,Kondo2015,Onoda2015}. 
The Ir conduction electron of Pr$_2$Ir$_2$O$_7$ is described 
by the Luttinger semimetal of this tight binding model.

Since all the pyrochlore iridates except Pr$_2$Ir$_2$O$_7$ 
experience a metal-insulator transition via the development 
of magnetic orders, a Hubbard-$U$ interaction is then 
introduced to capture this correlation driven Mott transition. 
As for Pr$_2$Ir$_2$O$_7$ that remains metallic, it is expected 
that the Hubbard-$U$ interaction merely renormalizes the bands 
but does not change the nature of the Luttinger semimetal. 
Without losing any generality, we set ${t_{\pi} = -2t_{\sigma}/3, 
t_{id} = -t_{\sigma}}$ throughout this work. 

Prior theoretical works, that focused on the Ir subsystem, have 
invoked the $k.p$ theory and the Luttinger model as the starting
point to analyze the correlation effect of the electrons~\cite{EGMoon,SavaryMoon,PhysRevB.95.085120,PhysRevB.92.035137,PhysRevB.95.075149,PhysRevB.93.205138,PhysRevLett.113.106401,PhysRevB.93.165109,PhysRevB.95.075101}. 
In our case, the ``Melko-Hertog-Gingras'' 
state of the Pr local moments has a large and finite ordering 
wavevector and necessarily connect the Ir bands near the $\Gamma$ point
with the bands near the ordering wavevector, so the lattice
effects cannot be ignored. As a result, we cannot start with the 
$k.p$ theory of the $\Gamma$ point at low energies, and instead, 
we should begin with the tight-binding model on the Ir pyrochlore 
lattice.

\subsection{Pr subsystem}

The Pr$^{3+}$ ion has a $4f^2$ electron configuration, and the $4f$ 
electron is well localized. The combination of the atomic spin-orbit 
coupling and the crystal electric field creates a two-fold degenerate 
ground state for the Pr$^{3+}$ ion. This two-fold ground state degeneracy 
defines the non-Kramers doublet nature of the Pr local moment, 
and a pseudospin-1/2 operator, $\boldsymbol{\tau}_i$, is introduced to 
operate on the two-fold degenerate ground states. The non-Kramers doublet  
has a peculiar property under the time reversal symmetry, {\sl i.e.} 
\begin{eqnarray}
\mathcal{T}: && \quad\quad  \tau^z_{i} \rightarrow - \tau^z_{i}, \\
\mathcal{T}: && \quad\quad  \tau^{x}_{i} \rightarrow + \tau^{x}_{i}, \\
\mathcal{T}: && \quad\quad  \tau^{y}_{i} \rightarrow + \tau^{y}_{i},
\end{eqnarray}
where the $z$ direction is defined locally on each sublattice and is 
given as the local (111) lattice direction of the pyrochlore system. 
Here, the magnetic dipolar moment is purely from the $\tau^z$ component,
and the transverse components are known to be the quadrupolar moments. 

Due to the spin-orbit-entangled nature of the Pr local moment, the 
effective interaction between the Pr local moments is anisotropic in 
the pseudospin space and also depends on the bond orientation. The 
general form of the interaction is~\cite{PhysRevB.78.094418,Onoda,SungbinBalents}
\begin{eqnarray}
\tilde{H}_{ex} &=& \sum_{ij} J_{z,ij}^{} \tau^z_i \tau^z_j 
          + \sum_{ij} J_{\perp,ij}^{} \sum_{\mu,\nu=x,y} 
                      \tau^{\mu}_i \tau^{\nu}_j,
\end{eqnarray}
where the interaction between the Ising component $\tau^z$ and the 
transverse component $\tau^{x,y}$ is strictly forbidden by time 
reversal symmetry. Here, $\tilde{H}_{ex}$ differs from $H_{ex}$ in 
Eq.~\eqref{eq1}. $\tilde{H}_{ex}$ contains all sources of interactions 
between the local moments, and is obtained by integrating out the Ir 
conduction electrons. $\tilde{H}_{ex}$ would contain both the RKKY 
interaction and $H_{ex}$. Since the Pr local moment is in the spin ice 
manifold, we thus expect the nearest-neighbor Ising interaction $J_{z,ij}$ 
is positive and dominant. The interaction between the transverse components 
creates the quantum fluctuation so that the system fluctuates quantum 
mechanically within the spin ice manifold. Clearly, the nearest-neighbor 
interaction alone cannot generate the finite momentum Ising order of 
the Pr system whose magnetic cell is twice the size of the crystal cell. 
Further neighbor interactions are required. We here introduce the third 
neighbor antiferromagnetic Ising interaction and approximate $\tilde{H}_{ex}$ as 
\begin{eqnarray}
\tilde{H}_{ex} \simeq 
\sum_{\langle ij \rangle} J^{}_{1z} \tau^z_i \tau^z_j +
\sum_{\langle\langle\langle ij \rangle\rangle\rangle}
J^{}_{3z}\tau^z_i \tau^z_j ,
\end{eqnarray}
where the interaction between the transverse components has been 
abandoned in this approximation. In our previous work that focuses 
on the quantum phase transition of the Pr subsystem, this quantum 
fluctuation is an important ingredient to understand the nature of 
the phase transition and the nearby phases. In contrast, our purpose 
in this paper is to understand the feedback effect on the Ir electron 
structure from the Pr Ising magnetic order, so the quantum dynamics of 
the Pr local moment is irrelevant for this purpose. In Sec.~\ref{sec3} 
of the paper, we would simply regard the Ising magnetic order 
that is observed in Pr$_2$Ir$_2$O$_7$ as a given condition,
and this exchange Hamiltonian is not invoked until in Sec.~\ref{sec4}
where the external Zeeman coupling competes with the exchange and
modifies the Pr magnetic order.  

The extended interaction for the Pr local moments in Pr$_2$Ir$_2$O$_7$
is expected because the RKKY interaction that is mediated by the 
Ir conduction electrons is not short-ranged. This is quite different    
from the usual rare-earth magnets where the exchange interaction is 
often short-ranged and mostly restricted to the nearest neighbors.
The long-range or extended RKKY interaction is the reason that we point 
out the Ir conduction drives the quantum phase transition of the Pr moments. 

Due to the Ising nature of the moment in the approximate exchange model, 
the ground state is antiferromagnetically ordered with an ordering 
wavevector ${{\boldsymbol Q}=2\pi(001)}$ for ${J_{3z} > 0}$. Clearly, 
the approximate model captures the observed magnetic order in Pr$_2$Ir$_2$O$_7$.

\subsection{Pr-Ir coupling}

\begin{figure}[t]
\centering
\includegraphics[width=8.5cm]{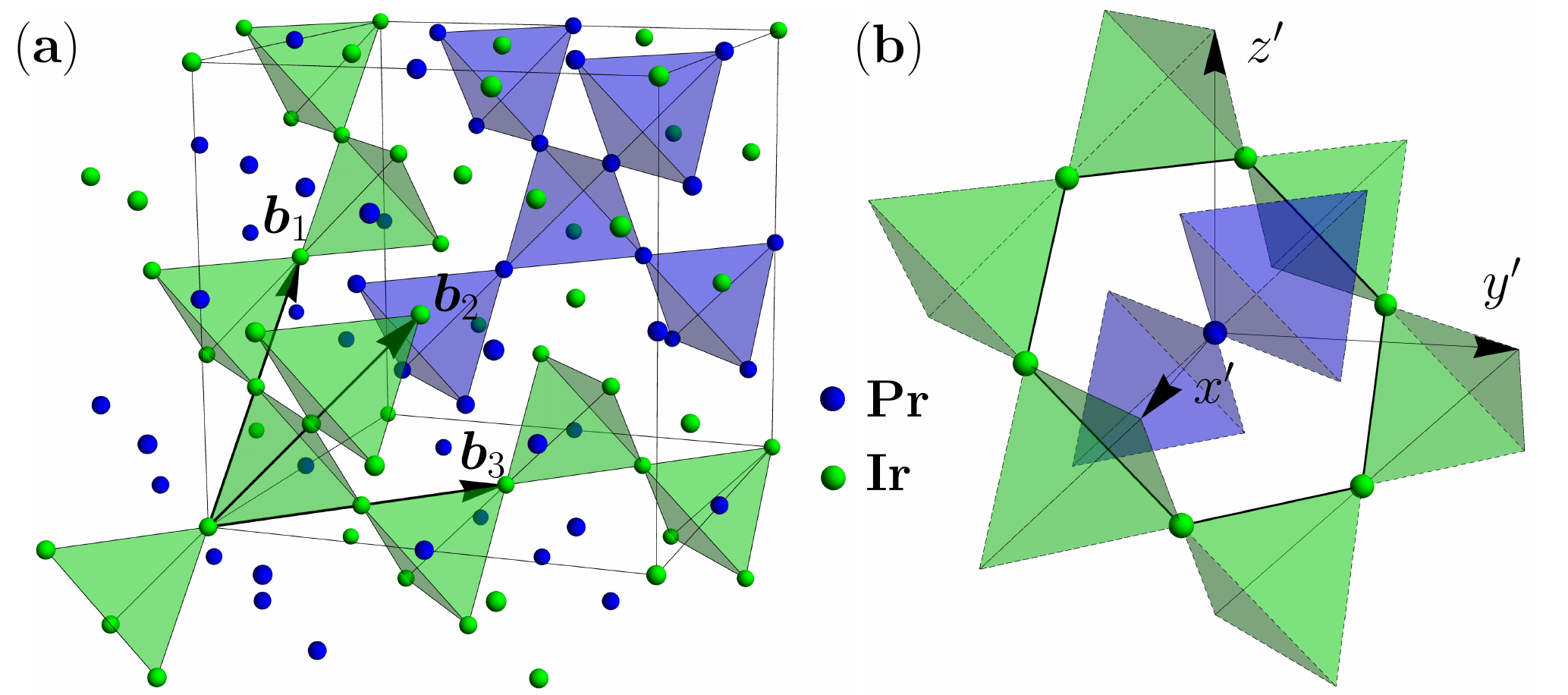}
\caption{The pyrochlore lattice structure for \ce{Pr2Ir2O7}. 
(a) Both Ir and Pr ions form pyrochlore lattices of corner-sharing tetrahedra. 
(b) For each Pr ion, six nearest Ir ions form a hexagon with the Pr ion 
in the center.}
\label{fig1} 
\end{figure}

Precisely because of the non-Kramers doublet nature of the Pr local moment,
it was pointed out in Ref.~\onlinecite{ChenHermele} based on the space group 
symmetry analysis that, the $\tau^z$ component couples to the spin density of 
the Ir conduction electron while the transverse component would couple to the 
electron density. The transverse component may also couple to 
the spin current that is even under time reversal~\cite{ChenHermele}. 
The general expression for the $f$-$d$ exchange between the Pr local moment
and the Ir spin density has been obtained in the previous work~\cite{ChenHermele}. 
The coupling between the transverse component $\tau^{x,y}$ and  
the Ir electron density was worked out in Ref.~\onlinecite{LeeArun}. 
Again, since it is the Ising component $\tau^z$ of the Pr local 
moment that develops the magnetic order in Pr$_2$Ir$_2$O$_7$,   
the leading order effect on the Ir conduction electron originates 
from the coupling between the Ir spin density and the Pr Ising component. 
Therefore, we consider the following $f$-$d$ exchange between the Pr 
Ising moment and the Ir spin density,~\cite{ChenHermele}
\begin{eqnarray}
H_{fd} =   \sum_{\langle ij \rangle} 
           \sum_{i\in\text{Pr}}
           \sum_{j \in \text{Ir}} 
           \tau^z_i \,
[ (d^\dagger_{j \alpha} \frac{\boldsymbol{\sigma}_{\alpha\beta}}{2} 
d^{}_{j\beta} ) \cdot {\boldsymbol v}_{ij} ],
\end{eqnarray}
where ${\boldsymbol v}_{ij}$ is a vector that defines the 
coupling between the Ir spin density and the Pr local moments.
For each Pr ion, there are six Ir ions nearby, and these six 
Ir ions form a hexagon with the Pr ion in the hexagon center 
(see Fig.~\ref{fig1}). Under the nearest-neighbor Kondo-like 
coupling approximation, the standard symmetry analysis gives 
for example
\begin{eqnarray}
        v_{11} & = & (0,0,0), \\
        v_{12} & = & (c_1,c_2,c_2), \\
        v_{13} & = & (c_2,c_1,c_2), \\
        v_{14} & = & (c_2,c_2,c_1), 
\end{eqnarray}
where $c_1, c_2$ are the two $f$-$d$ exchange parameters,
and other ${\boldsymbol v}_{ij}$'s can be obtained by 
simple lattice symmetry operations~\cite{ChenHermele}. 
The choices of the Pr and Ir sublattices are defined 
in Appendix.~\ref{ssec1}. 

\subsection{Zeeman coupling}

Finally, we introduce the Zeeman coupling. Because only the $\tau^z$ 
is odd under time reversal, we have the Zeeman coupling
\begin{eqnarray}
H_{\text{Zeeman}} & = & -  g\mu_{\text B} B \sum_{i \in \text{Pr}} 
\tau_i^{z} (\hat{z}_i \cdot \hat{n})
\nonumber \\
& \equiv & - h \sum_{i \in \text{Pr}} \tau_i^{z} 
(\hat{z}_i \cdot \hat{n}),
\end{eqnarray}
where $\hat{n}$ is the direction of the external magnetic field. 
The $\hat{z}_i$ direction is defined locally for each sublattice 
of the Pr subsystem.

\subsection{Energy scales}

Clearly, the largest energy scale in the model is the bandwidth and 
interaction of the Ir conduction interaction. The second largest energy 
scale is the $f$-$d$ exchange coupling. The lowest ones would be the 
exchange coupling between the Pr moments and the Zeeman coupling. 
Since the Zeeman coupling can be tuned experimentally, the magnetic
state of the Pr local moments can thus be manipulated by the external 
magnetic field.

\section{Dirac band touchings and Weyl nodes of the Iridium subsystem}
\label{sec3}

For \ce{Pr2Ir2O7}, the Ir conduction electrons were found to develop 
a Luttinger semimetallic band structure that is similar to the 
bulk HgTe~\cite{Kondo2015,PhysRevLett.106.126803,Bernevig1757,Onoda2015,Armitage2017}. 
It is well-known 
that the Luttinger semimetal is a parent state of various topological 
phases such as topological insulator and Weyl semimetal~\cite{Kondo2015,PhysRevLett.106.126803,Bernevig1757}. The Pr Ising order breaks the time reversal 
symmetry, and the time reversal symmetry breaking is transmitted to 
the Luttinger semimetal of the Ir subsystem through the $f$-$d$ exchange. 
We here study the band structure reconstruction of the Ir $5d$ electrons 
through the above mechanism.

\begin{figure}[t]
\centering
\includegraphics[width=8.7cm]{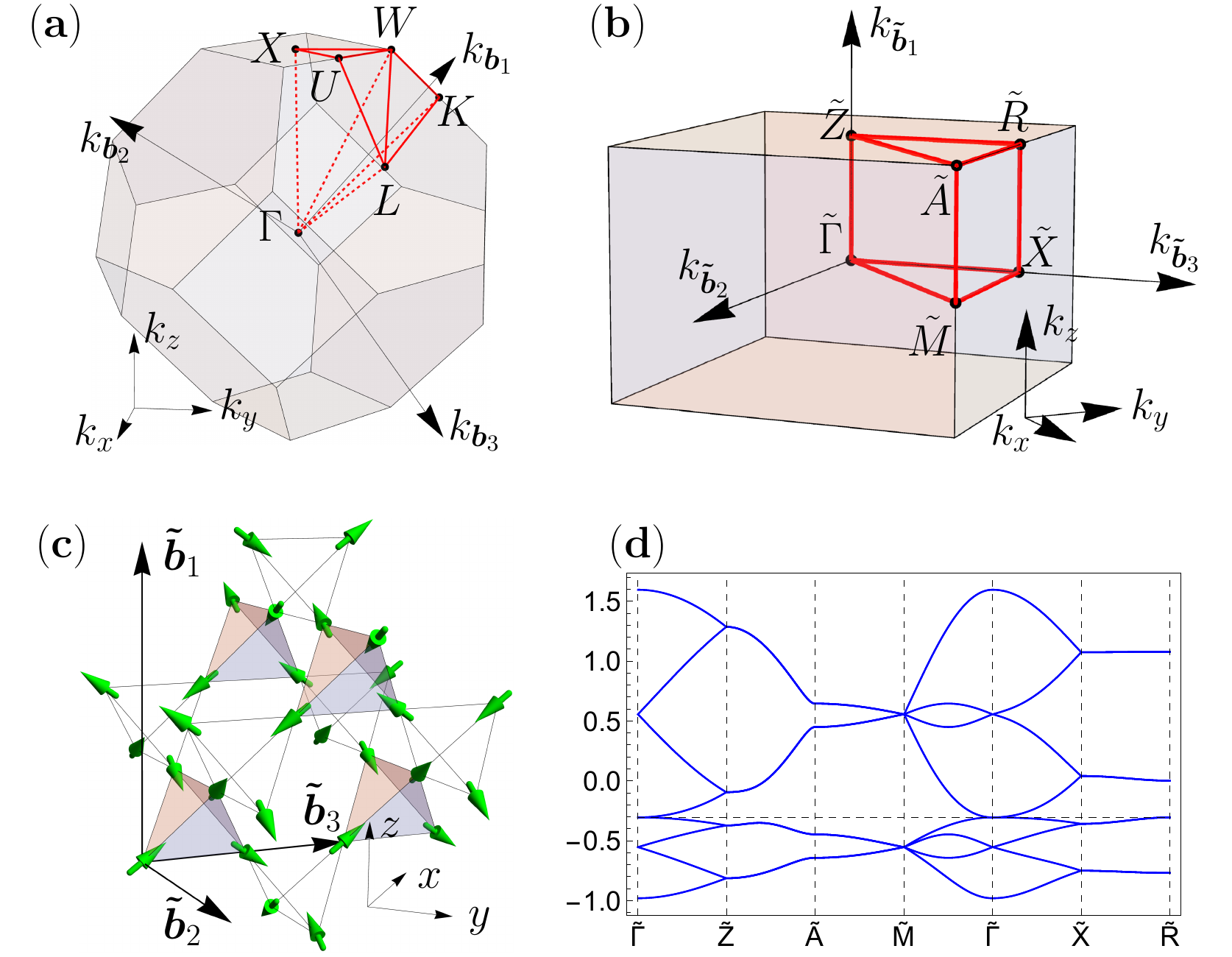}
\caption{
(a) The Brillouin zone of the original pyrochlore lattice. 
(b) Under the ${{\boldsymbol Q}=2\pi(001)}$ ``Melko-Hertog-Gingras'' spin 
ice state, the unit cell is enlarged. The plot is the magnetic Brillouin zone 
corresponding to the enlarged unit cell. (c) The spin configuration of the 
``Melko-Hertog-Gingras'' spin ice state. It is a 2-in 2-out spin wave with 
${{\boldsymbol Q}=2\pi(001)}$ ordering wavevector. 
(d) The folded energy band of the Ir conduction electrons without $f$-$d$ 
exchange term develops a quadratic touching at $\tilde{\Gamma}$ point. 
High symmetry momentum lines are defined in (b) as red lines.}
\label{fig2}
\end{figure}

\subsection{Emergent Dirac band touchings}

The Pr local moments were found to develop the ``Melko-Hertog-Gingras'' spin ice 
state in the recent samples with different Ir and O contents from the old ones. 
The ``Melko-Hertog-Gingras'' spin state breaks the time reversal and the lattice
translation by doubling the crystal unit cell. Due to this interesting magnetic 
ordering structure, the combination of the time reversal and certain lattice 
translations remains to be a symmetry of the system. As we show below, this 
symmetry leads to a remarkable band structure property of the Ir subsystem 
after the band reconstruction. 

\begin{figure}[t]
\centering
\includegraphics[width=0.48\textwidth]{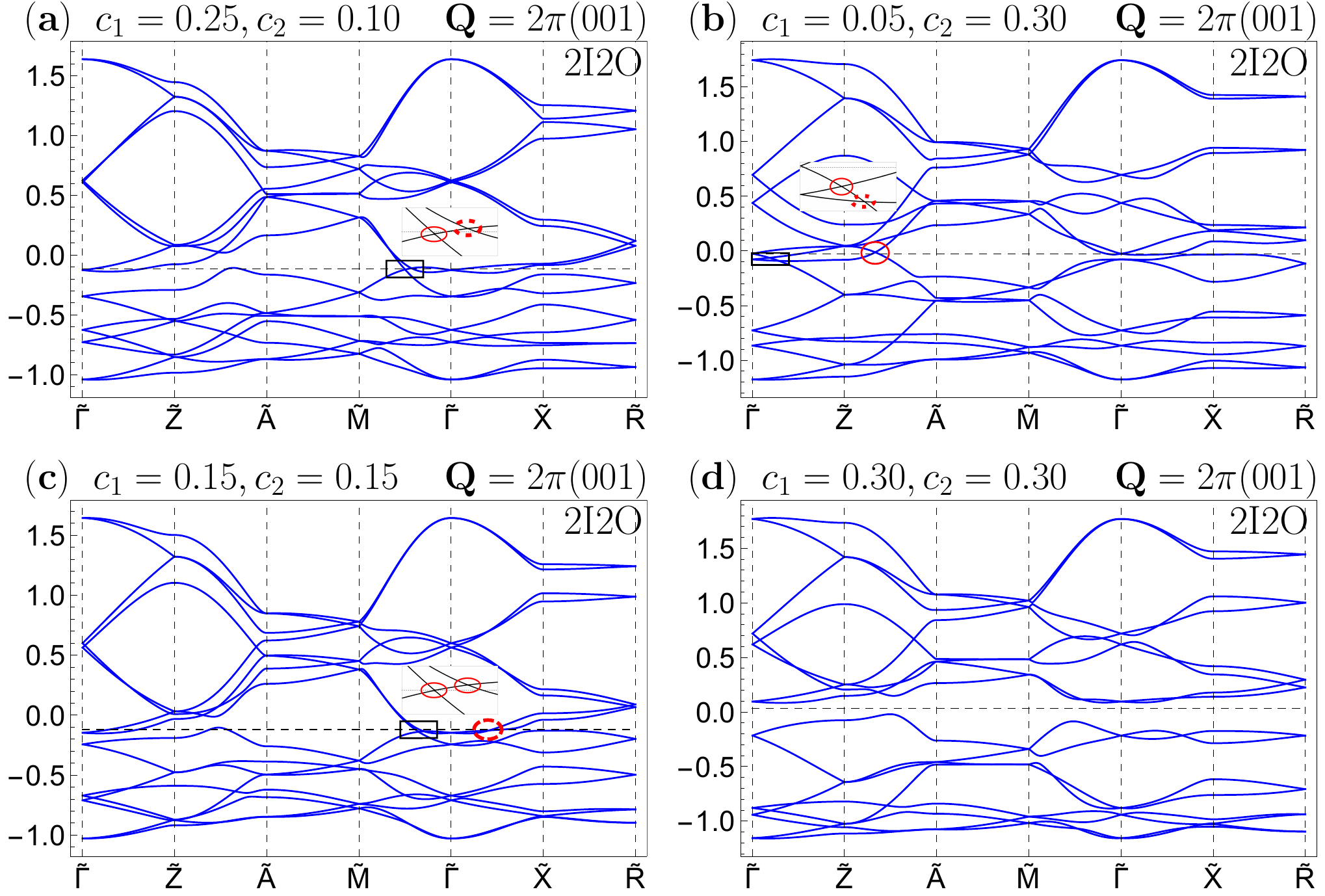}
\caption{Evolution of the Ir band structure as 
a function of $f$-$d$ exchange parameters $c_1$ and $c_2$. 
The dashed (solid) circle marks the usual (double) Weyl node.
Here, usual Weyl node has linear dispersions along all three
momentum directions while the double Weyl node has 
quadratic dispersions along two momentum direction 
and linear along one momentum direction~\cite{dWSM}. 
For sufficiently large parameters (d), the Weyl nodes 
disappear and a band gap is opened. Here, ``2I2O'' refers
to the ``2-in 2-out'' spin configuration. The energy unit 
in the plots is $t_{id}$. The dashed line refers to the Fermi energy.}
\label{fig3}
\end{figure}

The reconstructed band structure of the Ir conduction electrons 
is governed by the Ir tight binding model and the $f$-$d$ exchange, 
${H_{tb}+H_{fd}}$. As a comparison, we first evaluate the Ir 
band structure in the magnetic Brillouin zone corresponding 
to the doubled unit cell due to the Pr Ising magnetic order. 
As we depict in Fig.~\ref{fig2}, the Ir conduction electron 
bands form a Luttinger semimetal in the absence of the Pr
magnetic order and give a quadratic band touching at 
the $\tilde{\Gamma}$ point. Without losing any generality, 
in Fig.~\ref{fig2} we choose the ``Melko-Hertog-Gingras'' spin 
state of the Pr moments to have a propagating wavevector 
${{\boldsymbol Q} = 2\pi(001)}$ and the band structure in 
Fig.~\ref{fig2}(d) is plotted in the magnetic brioullin zone 
of Fig.~\ref{fig2}(b). Before the appearance of the Pr Ising  
order, the system has both time reversal ($\mathcal{T}$) and 
inversion ($\mathcal{I}$) symmetries, 
and each band of the Ir electrons has a two-fold Kramers degeneracy. 
The quadratic band touching at the $\tilde{\Gamma}$ point 
results from the cubic symmetry. As the Pr magnetic order 
appears, the Ir band structure is immediately modified. Before 
we present the reconstructed band structure in details, we 
first understand the band structure properties from the symmetry 
point of view. For our choice of the propagating wavevector, 
the ``Melko-Hertog-Gingras'' spin state breaks the lattice translations, 
$t_1$ and $t_2$. Here, $t_1$ and $t_2$ translate the system by 
the lattice basis vector ${{\boldsymbol b}_1 \equiv (0,1/2,1/2)}$ 
and ${{\boldsymbol b}_2 \equiv (1/2,0,1/2)}$, respectively.  
It turns out that, the combination of time reversal and 
$t_1$ or $t_2$, {\sl i.e.},
\begin{equation}
\tilde{\mathcal T}_1 \equiv  t_1 \circ {\mathcal T} , 
\quad 
\tilde{\mathcal T}_2 \equiv  t_2 \circ {\mathcal T} ,
\end{equation}
remains to be a symmetry of the system after the development of the Pr 
magnetic order. These two symmetries of the ``Melko-Hertog-Gingras'' spin
state are analogous to the staggered time reversal of the antiferromagnetic 
N\'{e}el state on a square lattice. Like the pure time reversal, 
$\tilde{\mathcal T}_1$ and $\tilde{\mathcal T}_2$ are anti-unitary operations.
Similar anti-unitary symmetry has been considered in the proposal of  
antiferromagnetc topological insulator by Mong, Essin and Moore~\cite{PhysRevB.81.245209}.
Due to the involvement of the lattice translations, $\tilde{\mathcal T}_1$ and 
$\tilde{\mathcal T}_2$ do not lead to the Kramers degeneracy for all the time 
reversal invariant momenta in the magnetic Brioullin zone. It is ready to confirm 
that, 
\begin{eqnarray}
&& \tilde{\mathcal T}_1 | \tilde{\Gamma}, \uparrow \rangle = i | \tilde{\Gamma}, \downarrow \rangle,
\quad\quad \tilde{\mathcal T}_2 | \tilde{\Gamma}, \uparrow \rangle =   i | \tilde{\Gamma}, \downarrow \rangle, \\
&& \tilde{\mathcal T}_1 | \tilde{M}, \uparrow \rangle =  i | \tilde{M}, \downarrow \rangle,
\quad\, \tilde{\mathcal T}_2 | \tilde{M} \uparrow \rangle =   -i | \tilde{M}, \downarrow \rangle ,\\
&& \tilde{\mathcal T}_1 | \tilde{R}, \uparrow \rangle = -i |  \tilde{R}, \downarrow \rangle,
\quad  \tilde{\mathcal T}_2 |  \tilde{R}, \uparrow \rangle =   -i |  \tilde{R}, \downarrow \rangle,
\end{eqnarray}
and ${\tilde{\mathcal T}_1^2 = \tilde{\mathcal T}_2^2 =-1}$ for the momentum points 
at $\tilde{\Gamma}$, $\tilde{M}$ and $\tilde{R}$;
and ${\tilde{\mathcal T}_1^2 = \tilde{\mathcal T}_2^2 =+1}$ for the momentum points 
at $\tilde{X}$, $\tilde{Z}$ and $\tilde{A}$. 
Note that $\tilde{\Gamma}$, $\tilde{M}$ and $\tilde{R}$ are also 
time reversal invariant momenta for the crystal Brioullin zone
while  $\tilde{X}$, $\tilde{Z}$ and $\tilde{A}$ are not. 
It immediately indicates that there 
are two-fold Kramers degeneracy at the $\tilde{\Gamma}$, $\tilde{M}$ and $\tilde{R}$ 
points, but not for the $\tilde{X}$, $\tilde{Z}$ and 
$\tilde{A}$ points. To confirm the above prediction, we carry out the explicit 
calculation of the Ir band structure in the presence of the Pr magnetic order. 
As we show in Fig.~\ref{fig3} for four specific choices of the $f$-$d$ exchange 
couplings, there exist emergent two-fold Kramers degeneracies with Dirac band 
touchings at the $\tilde{\Gamma}$, $\tilde{M}$ and $\tilde{R}$ points. 

\subsection{Magnetic Weyl nodes}

Besides the emergent and symmetry protected Dirac band touchings at the $\tilde{\Gamma}$, 
$\tilde{M}$ and $\tilde{R}$ points, we discover the presence of the Weyl nodes in the 
reconstructed Ir band structure in Fig.~\ref{fig3}. The reconstructed Ir band structure 
is determined by the $f$-$d$ exchange couplings. The actual couplings of the $f$-$d$ 
exchange in the material Pr$_2$Ir$_2$O$_7$ are unknown to us. To proceed, 
we fix the tight-binding part of the Ir hopping Hamiltonian and study the 
band structure phase diagram of Ir electrons by varying the $f$-$d$ 
exchange couplings. This approach is not designed to be self-consistent, but 
is phenomenological. The Pr Ising order, that is observed experimentally, 
is used as the input information to the Ir band structure calculation in this 
section. We expect, the realistic case for Pr$_2$Ir$_2$O$_7$ would be located 
at one specific parameter point in the phase diagram. It is possible that 
the pressure could vary the exchange couplings and allow the system to access 
different parameters of the phase diagram. 

\begin{figure}[t]
\centering
\includegraphics[width=7cm]{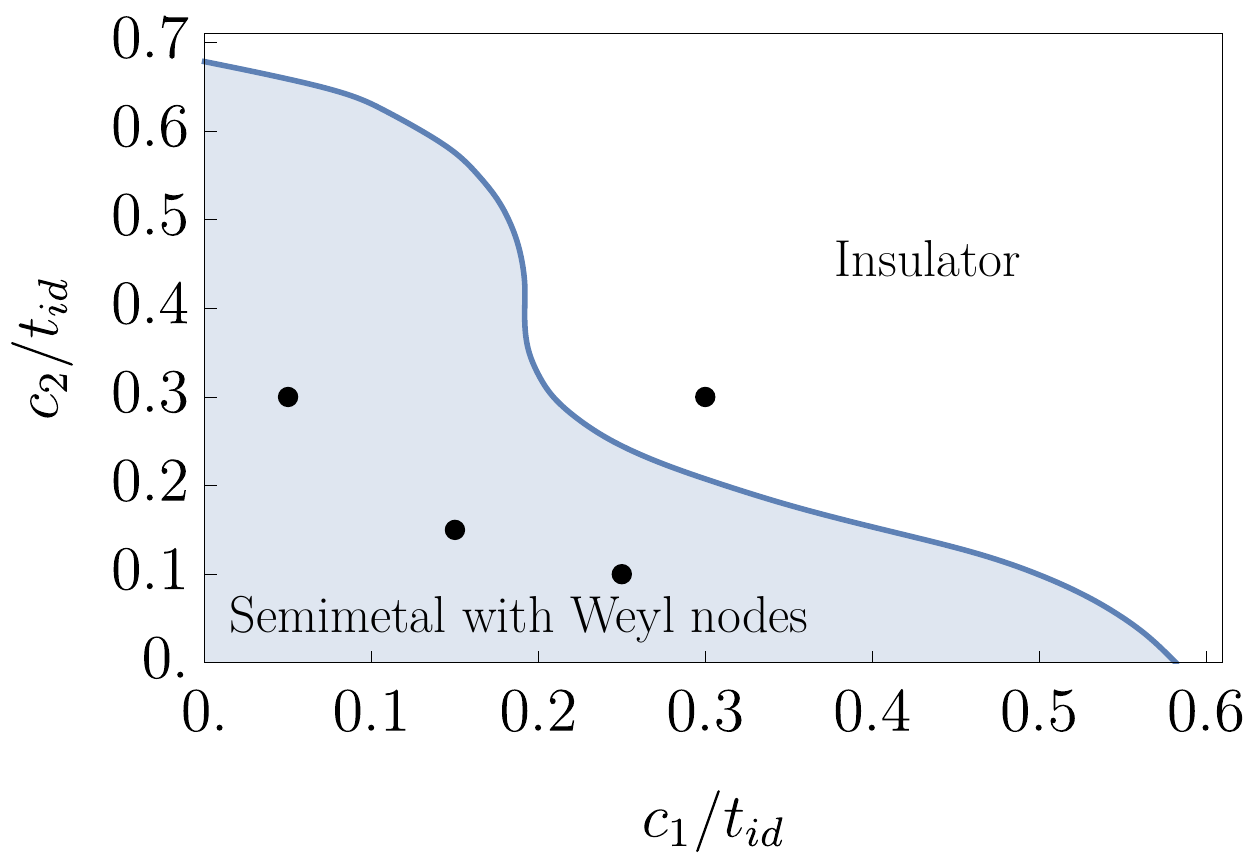}
\caption{Phase diagram for the Ir band structure in the 
parameter space of the $f$-$d$ exchange couplings.
The bold dots refer to the four parameter choices
in Fig.~\ref{fig3}. 
}
\label{fig4}
\end{figure}

In Fig.~\ref{fig4}, we depict our phase diagram according to the exchange couplings.
For small exchange couplings, a semimetal is always obtained. 
The name ``semimetal'' here not only refers to the Dirac band touching or dispersion
at some time reversal invariant momenta, but also refers to the (topologically protected) 
Weyl nodes in the magnetic Brioullin zone. In fact, Weyl semimetal with the surface Fermi 
arcs was first predicted for pyrochlore iridates with the all-in all-out magnetic order,
and the magnetic order is suggested to be driven by the Ir electron correlation~\cite{WanXG}. 
In our result here, the magnetic order comes from the Pr Ising order, and the time
reversal symmetry breaking is then transmitted to the Ir conduction electron via the $f$-$d$ 
exchange. The magnetic order is not the simple all-in all-out magnetic order. 
It was also suggested that the correlation-driven Weyl semimetal for pyrochlore 
iridates appears in a rather narrow parameter regime~\cite{YBKim1}. The $f$-$d$ exchange, 
however, could significantly enlarge the parameter regime for Weyl semimetal~\cite{ChenHermele}. 
Indeed, in Fig.~\ref{fig4}, the semimetal region does support several Weyl nodes 
near the Fermi level, and thus, we expect the usual properties for Weyl 
semimetal~\cite{WanXG} to hold in this regime. Moreover, since the $f$-$d$ 
exchange coupling is much smaller than the effective hoppings of the Ir electrons, 
so the realistic case for the ordered Pr$_2$Ir$_2$O$_7$ is expected to occur in 
the semimetallic phase of Fig.~\ref{fig4}. 

\section{Role of external magnetic fields}
\label{sec4}

To further control the property of the system, we suggest to apply
an (uniform) external magnetic field to the system. As we have explained
in Sec.~\ref{sec1}, the magnetic field would primarily couple to the Pr 
local moments. A uniform magnetic field induces a finite magnetic polarization
on the Pr local moments, and thus breaks the $\tilde{\mathcal T}_1$ and 
$\tilde{\mathcal T}_2$ symmetries of the ordered Pr$_2$Ir$_2$O$_7$. As a 
consequence, the emergent Dirac band touchings at the $\tilde{\Gamma}$, 
$\tilde{M}$ and $\tilde{R}$ points, that are protected by the 
$\tilde{\mathcal T}_1$ and $\tilde{\mathcal T}_2$ symmetries, 
should disappear immediately in a generic magnetic field along
a random direction. Here the choice of a random direction for 
the magnetic field simply avoids the accidental degeneracy/band touching 
that is protected by the reduced lattice symmetry of the system
if the field is applied along high symmetry directions.  

\begin{figure}[t]
\centering
\includegraphics[width=8.7cm]{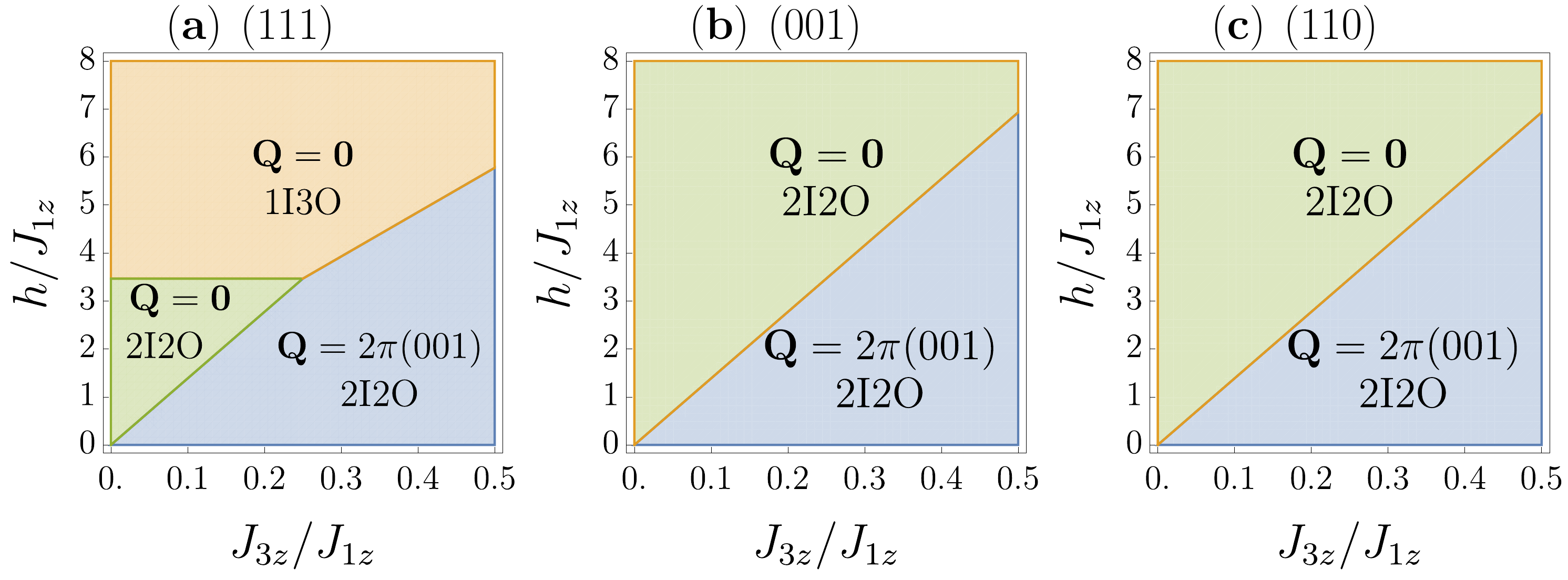}
\caption{Phase diagram of the Pr local moments under the 
external magnetic fields along different directions.
Here ``1I3O'' refers to ``1-in 3-out'' spin configuration. 
}
\label{fig5}
\end{figure}

Unlike the previous section where the Ir band structure is controlled by 
the $f$-$d$ exchange and the Ir tight-binding model, the Ir band structure 
in the magnetic field requires the knowledge of the Pr magnetic state 
that is modified by the external magnetic field. As we have explained 
in Sec.~\ref{sec1}, the external magnetic field first modifies the Pr 
magnetic state and then indirectly influences the Ir band structure 
through the $f$-$d$ exchange interaction. For the Pr subsystem, we 
consider the following Hamiltonian,
\begin{eqnarray}
H_{\text{Pr}} = \tilde{H}_{ex} + H_{\text{Zeeman}},
\end{eqnarray}
where the exchange part includes both the first neighbor and third neigbhor
Ising exchange interactions. Since here the Pr local moment is set to be an 
Ising degree of freedom, it is ready to obtain the magnetic phase diagram 
of the Pr moments by comparing energies of candidate ground states. 
The magnetic phase diagram for the Pr moments is depicted in Fig.~\ref{fig5},
where three different directions of magnetic fields are considered. 

\begin{figure}[t]
\centering
 \includegraphics[width=0.48\textwidth]{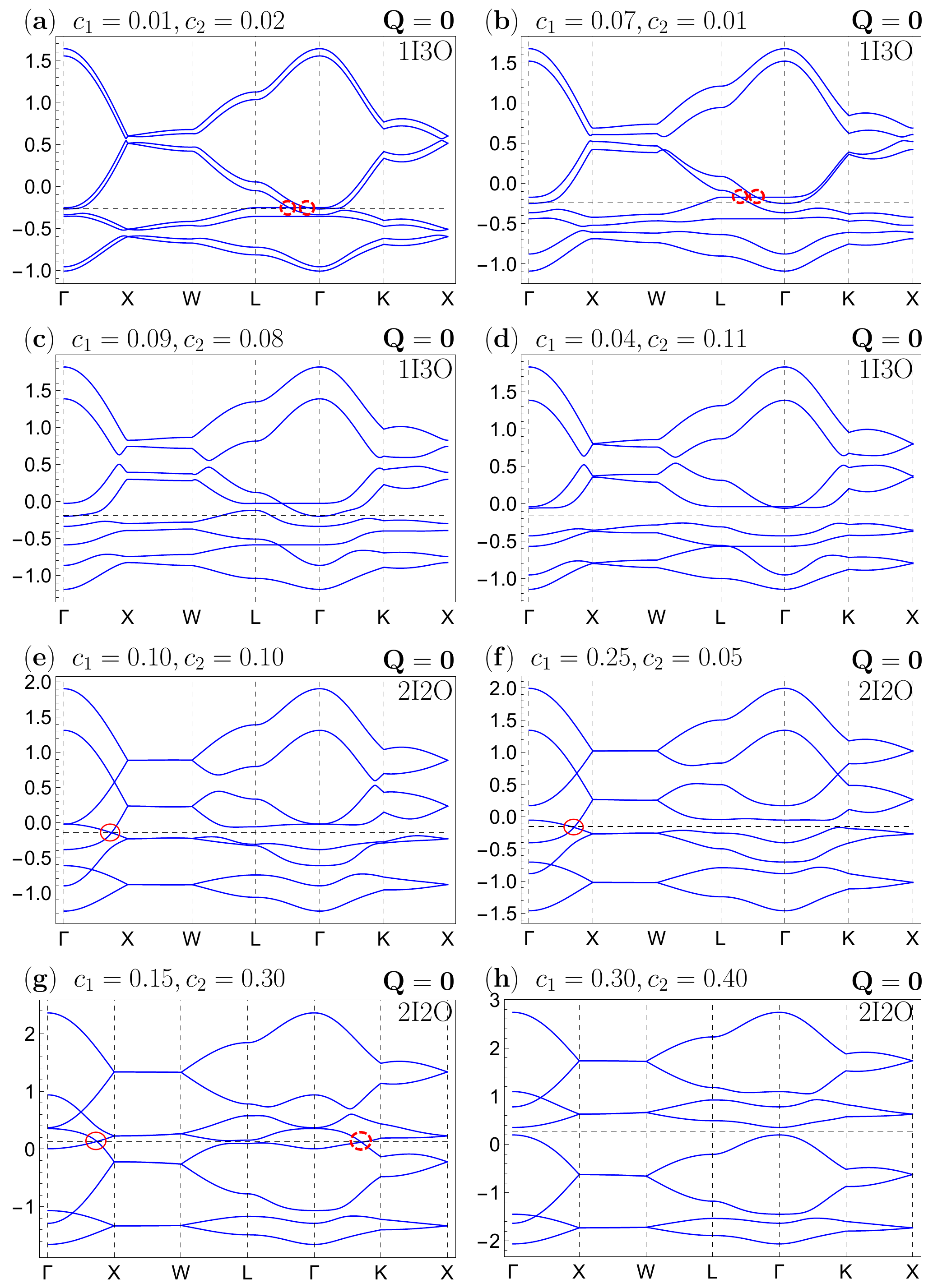}
\caption{Evolution of the Ir band structure as a function of $f$-$d$ 
exchange parameters for (a-d) ``1-in 3-out'' and (e-h) ``2-in 2-out'' 
Pr magnetic states with ${\boldsymbol{Q}}=0$ from Fig.~\ref{fig5}(a). 
The dashed (solid) circle marks the usual (double) Weyl node. 
 In (a) and (b), one band 
from $L$ to $\Gamma$ is flat. This is accidental for the nearest-neighbor 
hopping model and could be dispersive if further neighbor hoppings 
are included~\cite{YBKim1}. In (g), the Weyl nodes are actually at 
different energies. The energy unit in the plots is $t_{id}$.
The dashed line refers to the Fermi energy. 
}
\label{fig6}
\end{figure}

Here we focus on one specific field orientation, ${\hat{n}\equiv (1,1,1)/\sqrt{3}}$, 
and evaluate the feedback of the Pr magnetic state on the Ir conduction electrons. 
Besides the original ``Melko-Hertog-Gingras'' spin state, two additional spin states  
are obtained. While the Ir band structure in the presence of ``Melko-Hertog-Gingras'' 
spin state stays the same as the ones in Sec.~\ref{sec3} under this approximation, 
this should be the caveat of the approximation of the Pr moment as
the Ising spin that ignores the quantum nature of the Pr moments. 
In reality, the magnetic field would create a finite polarization   
for the Pr local moment and modifies the Ir band structure immediately, even
though the modification can be small. This would allow us to move the positions 
of the Weyl nodes in the momentum space. The other two spin configurations
of the Pr moments, that result from strong magnetic field, have an ordering 
wavevector ${\boldsymbol Q}={\boldsymbol 0}$ and restores the lattice 
translation symmetry. Hence, we expect two different Ir band structures 
for these spin configurations. In Fig.~\ref{fig6}, we depict the Ir band 
structures for specific choices of the $f$-$d$ exchanges with two 
${\boldsymbol Q}={\boldsymbol 0}$ spin configurations from the phase 
diagram in Fig.~\ref{fig5}(a). Our explicit calculation of the Ir band 
structure in Fig.~\ref{fig6} shows that the Dirac band touchings at the 
$\tilde{\Gamma} (\equiv \Gamma)$, $\tilde{R} (\equiv L)$ points are absent in the 
magnetic field, and now the magnetic unit cell is now identical to 
the crystal unit cell. Moreover, although the time reversal symmetry 
breaking is transmited by the Pr spin configuration due to the external
magnetic field, the overall effect is that one applies the time reversal 
symmetry breaking to the Ir Luttinger semimetal. Since Luttinger semimetal 
can be regarded as the parent state of the Weyl semimetal, it seems natural 
to expect the occurrence of the Weyl nodes. Indeed, as we show in 
Fig.~\ref{fig7}, we obtain the Weyl semimetal (or Weyl metal) for 
a large parameter regime in the phase diagram. 

With large magnetic fields along (001) and (110) directions, 
the ${\boldsymbol Q}={\boldsymbol 0}$ state is obtained for 
the Pr moments under this approximation (see Fig.~\ref{fig5}(b) 
and (c)). This Pr spin state is the same as one of the spin states 
when the field is applied along (111) direction, and there it does 
not bring different Ir band structures under this approximation.

\section{Discussion}
\label{sec5}

We here summarize our understanding of the rich physics in
Pr$_2$Ir$_2$O$_7$ and suggest future experiments to further 
reveal its physics. In the previous field theory work by one of the 
authors, we pointed out that the Pr subsystem of Pr$_2$Ir$_2$O$_7$ 
is proximate to a quantum phase transition from the $U(1)$ quantum 
spin liquid to the magnetic order. The proximate magnetic order, 
that is obtained from the condensation of the ``magnetic monopoles''
in the $U(1)$ quantum spin liquid, breaks the lattice translation
and is precisely the one that is observed in the neutron scattering 
experiments. This theoretical work indicates that the paramagnetic 
state of disordered Pr$_2$Ir$_2$O$_7$ sample is a $U(1)$ quantum spin 
liquid. In the current paper, we focus on the magnetically ordered 
Pr$_2$Ir$_2$O$_7$ sample. We have developed a systematic modelling 
to understand the interplay between the Ir conduction electrons 
and the Pr local moments for the material Pr$_2$Ir$_2$O$_7$. 
We use the existing experimental results, such as the Luttinger 
semimetal of the Ir conduction electrons and the 
``Melko-Hertog-Gingras'' spin state of the Pr local moments,
as the input information for our theoretical framework, 
and study the band reconstruction of the Ir conduction
electrons in the presence of the Pr magnetic order. We predict
that the symmetry protected Dirac cones emerge at part of the 
time reversal invariant momenta and the symmetry protection
comes from the magnetic translation symmetry of the 
``Melko-Hertog-Gingras'' spin state for the Pr subsystem. 
Moreover, there generically exist Weyl nodes of different
kinds both in the ordered Pr$_2$Ir$_2$O$_7$ samples and 
Pr$_2$Ir$_2$O$_7$ in the external magnetic fields.

\begin{figure}[t]
\centering
\includegraphics[width=8.5cm]{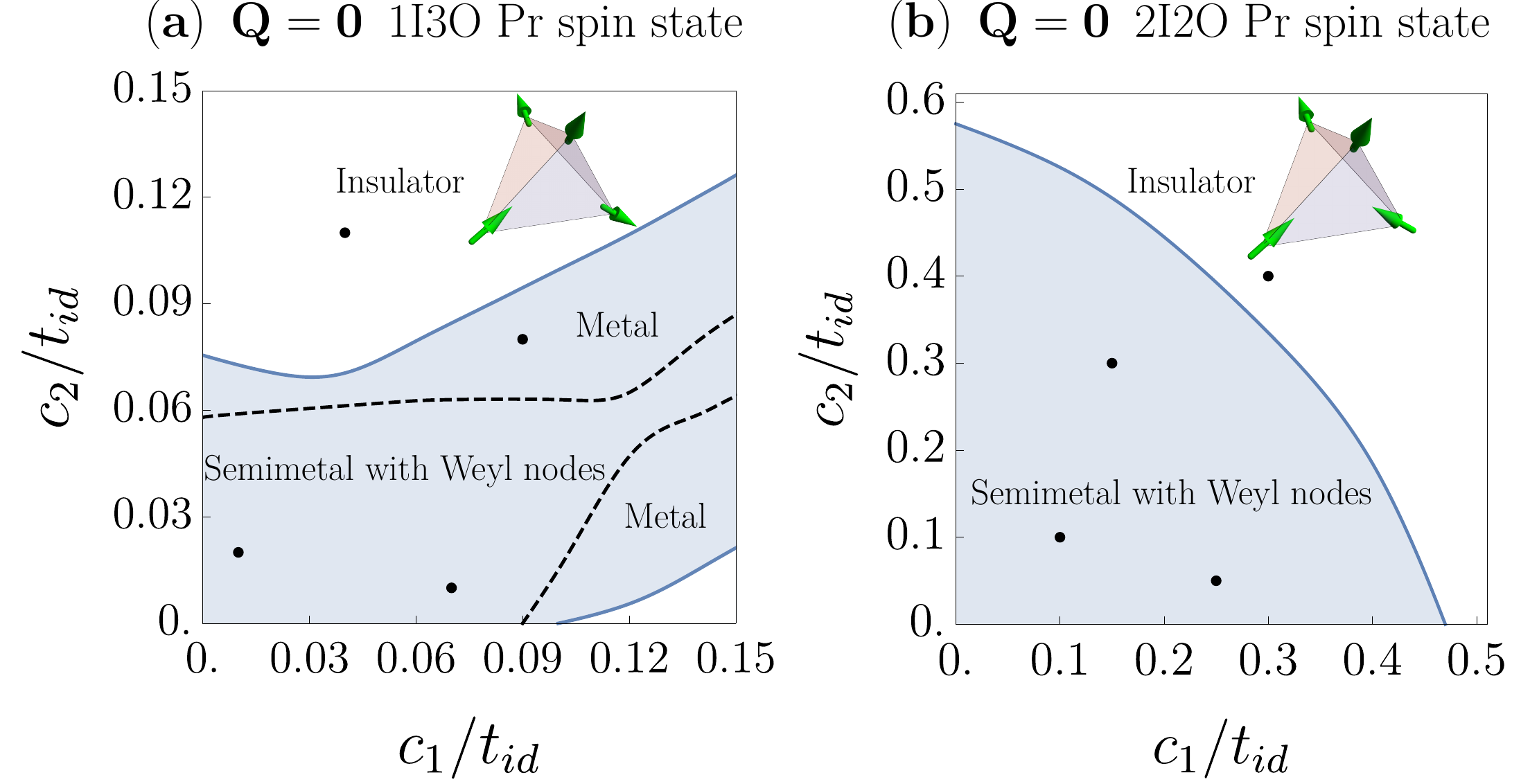}
\caption{Phase diagram for the Ir band structure in the 
parameter space of the $f$-$d$ exchange couplings for different 
Pr magnetic states from Fig.~\ref{fig5}(a). 
The bold dots in the plots are the parameter choices for Fig.~\ref{fig6}.}
\label{fig7}
\end{figure}

Based on our prediction about the non-trivial Ir band structure
after the reconstruction from the Pr magnetic state, we here propose
the future experiments. Certainly, the non-trivial features of 
the Ir band structure in the ordered Pr$_2$Ir$_2$O$_7$ sample
(without the magnetic field) would be best detected by the 
angle-resolved photo-emission spectroscopy (ARPES). The optical 
measurements can also be useful for the inter-band particle-hole 
transition near the band touching points. The Dirac band touchings
at some of the time reversal invariant momenta, that are protected 
by the magnetic translation of the ``Melko-Hertog-Gingras'' spin 
state, would immediately disappear when the magnetic field is applied. 
This prediction may be a sharp feature for the experimental 
confirmation. Besides the direct band structure measurement, 
the magneto-transport can be a useful probe. Due to the breaking 
of the cubic symmetry, the Weyl semimetal that is induced by the 
external magnetic field would show anomalous Hall effect. 
Finally, we point out the field-driven metal-insulator transition.
Although it was not emphasized in Sec.~\ref{sec4}, 
the large portion of the semimetal region in the phase diagram 
of Fig.~\ref{fig4} is converted into the insulating region 
in the phase diagram of Fig.~\ref{fig7}(a). 
From the experience in Nd$_2$Ir$_2$O$_7$ with the dipole-octupole 
Nd$^{3+}$ magnetic ions~\cite{Nd2Ir2O7,PhysRevLett.112.167203,PhysRevB.94.201114,ChenHermele},
this field-driven metal-insulator transition via the $f$-$d$ exchange 
could be the most visible experimental signature in the transport 
measurement.

\section{Acknowledgments}
    
We are indebted to D.-H. Lee and F.-C. Zhang for their advice that 
wakes me up to write out and/or submit our papers including this 
one here. We acknowledge C. Broholm for the early discussion about 
the Pr magnetic order of \ce{Pr2Ir2O7} at University of Cambridge 
in the summer of 2015, a correpondence with S. Nakatsuji, useful 
conversation with Roger Mong, Zhong Wang and Yi Zhou, and an early 
collaboration with Michael Hermele from CU Boulder. This work is 
supported by the ministry of science and technology of China with the 
grant no.2016YFA0301001, the start-up fund and the first-class university 
construction fund of Fudan University, and the thousand-youth-talent 
program of China.

\appendix

\section{Sublattices and crystal momenta for \ce{Pr2Ir2O7}}
\label{ssec1}

In \ce{Pr2Ir2O7}, both Ir and Pr pyrochlore lattices are 
composed of linked tetrahedra and can be viewed as FCC lattice 
with primitive lattice vectors 
    \begin{eqnarray}
        \bm{b}_1 & = & (0,\frac{1}{2},\frac{1}{2}), \\
        \bm{b}_2 & = & (\frac{1}{2},0,\frac{1}{2}), \\
        \bm{b}_3 & = & (\frac{1}{2},\frac{1}{2},0).
    \end{eqnarray}
After choosing one of the Ir site as reference point, the 
reference positions of four Ir sublattices can be set to be 
\begin{eqnarray}
        \text{Ir}_1 & = & (0,0,0),\quad
        \text{Ir}_2  =  (0,\frac{1}{4},\frac{1}{4}), \\
        \text{Ir}_3 & = & (\frac{1}{4},0,\frac{1}{4}),\quad
        \text{Ir}_4  =  (\frac{1}{4},\frac{1}{4},0).
\end{eqnarray}
Likewise, for the Pr subsystem we have the reference 
positions of four Pr sublattices as
\begin{eqnarray}
        \text{Pr}_1 & = & (0,\frac{1}{2},0), \quad
        \text{Pr}_2  =  (0,\frac{3}{4},\frac{1}{4}), \\
        \text{Pr}_3 & = & (\frac{1}{4},\frac{1}{2},\frac{1}{4}), \quad
        \text{Pr}_4  =  (\frac{1}{4},\frac{3}{4},0).
\end{eqnarray}

The crystal momenta in Fig.~\ref{fig2} are 
\begin{eqnarray}
\tilde{\Gamma} &=& (0,0,0), 
\tilde{M} = (2\pi, 0,0),   
\tilde{R} = (\pi,\pi,\pi), \\
\tilde{X} &=& (\pi,\pi,0),   
\tilde{Z} = (0,0,\pi), \,\,\,
\tilde{A} = (2\pi,0,\pi). 
\end{eqnarray}

\bibliography{refs}

\begin{thebibliography}{59}%
\makeatletter
\providecommand \@ifxundefined [1]{%
 \@ifx{#1\undefined}
}%
\providecommand \@ifnum [1]{%
 \ifnum #1\expandafter \@firstoftwo
 \else \expandafter \@secondoftwo
 \fi
}%
\providecommand \@ifx [1]{%
 \ifx #1\expandafter \@firstoftwo
 \else \expandafter \@secondoftwo
 \fi
}%
\providecommand \natexlab [1]{#1}%
\providecommand \enquote  [1]{``#1''}%
\providecommand \bibnamefont  [1]{#1}%
\providecommand \bibfnamefont [1]{#1}%
\providecommand \citenamefont [1]{#1}%
\providecommand \href@noop [0]{\@secondoftwo}%
\providecommand \href [0]{\begingroup \@sanitize@url \@href}%
\providecommand \@href[1]{\@@startlink{#1}\@@href}%
\providecommand \@@href[1]{\endgroup#1\@@endlink}%
\providecommand \@sanitize@url [0]{\catcode `\\12\catcode `\$12\catcode
  `\&12\catcode `\#12\catcode `\^12\catcode `\_12\catcode `\%12\relax}%
\providecommand \@@startlink[1]{}%
\providecommand \@@endlink[0]{}%
\providecommand \url  [0]{\begingroup\@sanitize@url \@url }%
\providecommand \@url [1]{\endgroup\@href {#1}{\urlprefix }}%
\providecommand \urlprefix  [0]{URL }%
\providecommand \Eprint [0]{\href }%
\providecommand \doibase [0]{http://dx.doi.org/}%
\providecommand \selectlanguage [0]{\@gobble}%
\providecommand \bibinfo  [0]{\@secondoftwo}%
\providecommand \bibfield  [0]{\@secondoftwo}%
\providecommand \translation [1]{[#1]}%
\providecommand \BibitemOpen [0]{}%
\providecommand \bibitemStop [0]{}%
\providecommand \bibitemNoStop [0]{.\EOS\space}%
\providecommand \EOS [0]{\spacefactor3000\relax}%
\providecommand \BibitemShut  [1]{\csname bibitem#1\endcsname}%
\let\auto@bib@innerbib\@empty
\bibitem [{\citenamefont {Hasan}\ and\ \citenamefont
  {Kane}(2010)}]{KaneReview}%
  \BibitemOpen
  \bibfield  {author} {\bibinfo {author} {\bibfnamefont {M.~Z.}\ \bibnamefont
  {Hasan}}\ and\ \bibinfo {author} {\bibfnamefont {C.~L.}\ \bibnamefont
  {Kane}},\ }\bibfield  {title} {\enquote {\bibinfo {title} {{Colloquium:
  Topological insulators}},}\ }\href {\doibase 10.1103/RevModPhys.82.3045}
  {\bibfield  {journal} {\bibinfo  {journal} {Rev. Mod. Phys.}\ }\textbf
  {\bibinfo {volume} {82}},\ \bibinfo {pages} {3045--3067} (\bibinfo {year}
  {2010})}\BibitemShut {NoStop}%
\bibitem [{\citenamefont {Qi}\ and\ \citenamefont
  {Zhang}(2011)}]{ShouchengReview}%
  \BibitemOpen
  \bibfield  {author} {\bibinfo {author} {\bibfnamefont {Xiao-Liang}\
  \bibnamefont {Qi}}\ and\ \bibinfo {author} {\bibfnamefont {Shou-Cheng}\
  \bibnamefont {Zhang}},\ }\bibfield  {title} {\enquote {\bibinfo {title}
  {Topological insulators and superconductors},}\ }\href {\doibase
  10.1103/RevModPhys.83.1057} {\bibfield  {journal} {\bibinfo  {journal} {Rev.
  Mod. Phys.}\ }\textbf {\bibinfo {volume} {83}},\ \bibinfo {pages}
  {1057--1110} (\bibinfo {year} {2011})}\BibitemShut {NoStop}%
\bibitem [{\citenamefont {Armitage}\ \emph {et~al.}()\citenamefont {Armitage},
  \citenamefont {Mele},\ and\ \citenamefont {Vishwanath}}]{VishwanathReview}%
  \BibitemOpen
  \bibfield  {author} {\bibinfo {author} {\bibfnamefont {N.P.}\ \bibnamefont
  {Armitage}}, \bibinfo {author} {\bibfnamefont {E.J.}\ \bibnamefont {Mele}}, \
  and\ \bibinfo {author} {\bibfnamefont {A.}~\bibnamefont {Vishwanath}},\
  }\bibfield  {title} {\enquote {\bibinfo {title} {{Weyl and Dirac Semimetals
  in Three Dimensional Solids}},}\ }\href@noop {} {\bibfield  {journal}
  {\bibinfo  {journal} {arXiv}\ }\textbf {\bibinfo {volume}
  {1705.01111}}}\BibitemShut {NoStop}%
\bibitem [{\citenamefont {Ando}\ and\ \citenamefont {Fu}(2015)}]{FuTCI}%
  \BibitemOpen
  \bibfield  {author} {\bibinfo {author} {\bibfnamefont {Yoichi}\ \bibnamefont
  {Ando}}\ and\ \bibinfo {author} {\bibfnamefont {Liang}\ \bibnamefont {Fu}},\
  }\bibfield  {title} {\enquote {\bibinfo {title} {{Topological Crystalline
  Insulators and Topological Superconductors: From Concepts to Materials}},}\
  }\href {\doibase 10.1146/annurev-conmatphys-031214-014501} {\bibfield
  {journal} {\bibinfo  {journal} {Annual Review of Condensed Matter Physics}\
  }\textbf {\bibinfo {volume} {6}},\ \bibinfo {pages} {361--381} (\bibinfo
  {year} {2015})}\BibitemShut {NoStop}%
\bibitem [{\citenamefont {Yu}\ \emph {et~al.}(2010)\citenamefont {Yu},
  \citenamefont {Zhang}, \citenamefont {Zhang}, \citenamefont {Zhang},
  \citenamefont {Dai},\ and\ \citenamefont {Fang}}]{Yu61}%
  \BibitemOpen
  \bibfield  {author} {\bibinfo {author} {\bibfnamefont {Rui}\ \bibnamefont
  {Yu}}, \bibinfo {author} {\bibfnamefont {Wei}\ \bibnamefont {Zhang}},
  \bibinfo {author} {\bibfnamefont {Hai-Jun}\ \bibnamefont {Zhang}}, \bibinfo
  {author} {\bibfnamefont {Shou-Cheng}\ \bibnamefont {Zhang}}, \bibinfo
  {author} {\bibfnamefont {Xi}~\bibnamefont {Dai}}, \ and\ \bibinfo {author}
  {\bibfnamefont {Zhong}\ \bibnamefont {Fang}},\ }\bibfield  {title} {\enquote
  {\bibinfo {title} {{Quantized Anomalous Hall Effect in Magnetic Topological
  Insulators}},}\ }\href {\doibase 10.1126/science.1187485} {\bibfield
  {journal} {\bibinfo  {journal} {Science}\ }\textbf {\bibinfo {volume}
  {329}},\ \bibinfo {pages} {61--64} (\bibinfo {year} {2010})}\BibitemShut
  {NoStop}%
\bibitem [{\citenamefont {Chang}\ \emph {et~al.}(2013)\citenamefont {Chang},
  \citenamefont {Zhang}, \citenamefont {Feng}, \citenamefont {Shen},
  \citenamefont {Zhang}, \citenamefont {Guo}, \citenamefont {Li}, \citenamefont
  {Ou}, \citenamefont {Wei}, \citenamefont {Wang}, \citenamefont {Ji},
  \citenamefont {Feng}, \citenamefont {Ji}, \citenamefont {Chen}, \citenamefont
  {Jia}, \citenamefont {Dai}, \citenamefont {Fang}, \citenamefont {Zhang},
  \citenamefont {He}, \citenamefont {Wang}, \citenamefont {Lu}, \citenamefont
  {Ma},\ and\ \citenamefont {Xue}}]{Chang167}%
  \BibitemOpen
  \bibfield  {author} {\bibinfo {author} {\bibfnamefont {Cui-Zu}\ \bibnamefont
  {Chang}}, \bibinfo {author} {\bibfnamefont {Jinsong}\ \bibnamefont {Zhang}},
  \bibinfo {author} {\bibfnamefont {Xiao}\ \bibnamefont {Feng}}, \bibinfo
  {author} {\bibfnamefont {Jie}\ \bibnamefont {Shen}}, \bibinfo {author}
  {\bibfnamefont {Zuocheng}\ \bibnamefont {Zhang}}, \bibinfo {author}
  {\bibfnamefont {Minghua}\ \bibnamefont {Guo}}, \bibinfo {author}
  {\bibfnamefont {Kang}\ \bibnamefont {Li}}, \bibinfo {author} {\bibfnamefont
  {Yunbo}\ \bibnamefont {Ou}}, \bibinfo {author} {\bibfnamefont {Pang}\
  \bibnamefont {Wei}}, \bibinfo {author} {\bibfnamefont {Li-Li}\ \bibnamefont
  {Wang}}, \bibinfo {author} {\bibfnamefont {Zhong-Qing}\ \bibnamefont {Ji}},
  \bibinfo {author} {\bibfnamefont {Yang}\ \bibnamefont {Feng}}, \bibinfo
  {author} {\bibfnamefont {Shuaihua}\ \bibnamefont {Ji}}, \bibinfo {author}
  {\bibfnamefont {Xi}~\bibnamefont {Chen}}, \bibinfo {author} {\bibfnamefont
  {Jinfeng}\ \bibnamefont {Jia}}, \bibinfo {author} {\bibfnamefont
  {Xi}~\bibnamefont {Dai}}, \bibinfo {author} {\bibfnamefont {Zhong}\
  \bibnamefont {Fang}}, \bibinfo {author} {\bibfnamefont {Shou-Cheng}\
  \bibnamefont {Zhang}}, \bibinfo {author} {\bibfnamefont {Ke}~\bibnamefont
  {He}}, \bibinfo {author} {\bibfnamefont {Yayu}\ \bibnamefont {Wang}},
  \bibinfo {author} {\bibfnamefont {Li}~\bibnamefont {Lu}}, \bibinfo {author}
  {\bibfnamefont {Xu-Cun}\ \bibnamefont {Ma}}, \ and\ \bibinfo {author}
  {\bibfnamefont {Qi-Kun}\ \bibnamefont {Xue}},\ }\bibfield  {title} {\enquote
  {\bibinfo {title} {{Experimental Observation of the Quantum Anomalous Hall
  Effect in a Magnetic Topological Insulator}},}\ }\href {\doibase
  10.1126/science.1234414} {\bibfield  {journal} {\bibinfo  {journal}
  {Science}\ }\textbf {\bibinfo {volume} {340}},\ \bibinfo {pages} {167--170}
  (\bibinfo {year} {2013})}\BibitemShut {NoStop}%
\bibitem [{\citenamefont {Liu}\ \emph {et~al.}(2016)\citenamefont {Liu},
  \citenamefont {Zhang},\ and\ \citenamefont {Qi}}]{QAHE}%
  \BibitemOpen
  \bibfield  {author} {\bibinfo {author} {\bibfnamefont {Chao-Xing}\
  \bibnamefont {Liu}}, \bibinfo {author} {\bibfnamefont {Shou-Cheng}\
  \bibnamefont {Zhang}}, \ and\ \bibinfo {author} {\bibfnamefont {Xiao-Liang}\
  \bibnamefont {Qi}},\ }\bibfield  {title} {\enquote {\bibinfo {title} {{The
  Quantum Anomalous Hall Effect: Theory and Experiment}},}\ }\href {\doibase
  10.1146/annurev-conmatphys-031115-011417} {\bibfield  {journal} {\bibinfo
  {journal} {Annual Review of Condensed Matter Physics}\ }\textbf {\bibinfo
  {volume} {7}},\ \bibinfo {pages} {301--321} (\bibinfo {year}
  {2016})}\BibitemShut {NoStop}%
\bibitem [{\citenamefont {Pesin}\ and\ \citenamefont
  {Balents}(2010)}]{PesinBalents}%
  \BibitemOpen
  \bibfield  {author} {\bibinfo {author} {\bibfnamefont {D.}~\bibnamefont
  {Pesin}}\ and\ \bibinfo {author} {\bibfnamefont {L.}~\bibnamefont
  {Balents}},\ }\bibfield  {title} {\enquote {\bibinfo {title} {Mott physics
  and band topology in materials with strong spin–orbit interaction},}\
  }\href {\doibase 10.1038/nphys1606} {\bibfield  {journal} {\bibinfo
  {journal} {Nature Physics}\ }\textbf {\bibinfo {volume} {6}},\ \bibinfo
  {pages} {376–381} (\bibinfo {year} {2010})}\BibitemShut {NoStop}%
\bibitem [{\citenamefont {Yanagishima}\ and\ \citenamefont
  {Maeno}(2001)}]{JPSJ.70.2880}%
  \BibitemOpen
  \bibfield  {author} {\bibinfo {author} {\bibfnamefont {Daiki}\ \bibnamefont
  {Yanagishima}}\ and\ \bibinfo {author} {\bibfnamefont {Yoshiteru}\
  \bibnamefont {Maeno}},\ }\bibfield  {title} {\enquote {\bibinfo {title}
  {Metal-nonmetal changeover in pyrochlore iridates},}\ }\href {\doibase
  10.1143/JPSJ.70.2880} {\bibfield  {journal} {\bibinfo  {journal} {Journal of
  the Physical Society of Japan}\ }\textbf {\bibinfo {volume} {70}},\ \bibinfo
  {pages} {2880--2883} (\bibinfo {year} {2001})}\BibitemShut {NoStop}%
\bibitem [{\citenamefont {Matsuhira}\ \emph {et~al.}(2011)\citenamefont
  {Matsuhira}, \citenamefont {Wakeshima}, \citenamefont {Hinatsu},\ and\
  \citenamefont {Takagi}}]{Pyrochloreiridate2002}%
  \BibitemOpen
  \bibfield  {author} {\bibinfo {author} {\bibfnamefont {Kazuyuki}\
  \bibnamefont {Matsuhira}}, \bibinfo {author} {\bibfnamefont {Makoto}\
  \bibnamefont {Wakeshima}}, \bibinfo {author} {\bibfnamefont {Yukio}\
  \bibnamefont {Hinatsu}}, \ and\ \bibinfo {author} {\bibfnamefont {Seishi}\
  \bibnamefont {Takagi}},\ }\bibfield  {title} {\enquote {\bibinfo {title}
  {{Metal–Insulator Transitions in Pyrochlore Oxides Ln$_2$Ir$_2$O$_7$}},}\
  }\href {\doibase 10.1143/JPSJ.80.094701} {\bibfield  {journal} {\bibinfo
  {journal} {Journal of the Physical Society of Japan}\ }\textbf {\bibinfo
  {volume} {80}},\ \bibinfo {pages} {094701} (\bibinfo {year}
  {2011})}\BibitemShut {NoStop}%
\bibitem [{\citenamefont {Wan}\ \emph {et~al.}(2011)\citenamefont {Wan},
  \citenamefont {Turner}, \citenamefont {Vishwanath},\ and\ \citenamefont
  {Savrasov}}]{WanXG}%
  \BibitemOpen
  \bibfield  {author} {\bibinfo {author} {\bibfnamefont {Xiangang}\
  \bibnamefont {Wan}}, \bibinfo {author} {\bibfnamefont {Ari~M.}\ \bibnamefont
  {Turner}}, \bibinfo {author} {\bibfnamefont {Ashvin}\ \bibnamefont
  {Vishwanath}}, \ and\ \bibinfo {author} {\bibfnamefont {Sergey~Y.}\
  \bibnamefont {Savrasov}},\ }\bibfield  {title} {\enquote {\bibinfo {title}
  {Topological semimetal and fermi-arc surface states in the electronic
  structure of pyrochlore iridates},}\ }\href {\doibase
  10.1103/PhysRevB.83.205101} {\bibfield  {journal} {\bibinfo  {journal} {Phys.
  Rev. B}\ }\textbf {\bibinfo {volume} {83}},\ \bibinfo {pages} {205101}
  (\bibinfo {year} {2011})}\BibitemShut {NoStop}%
\bibitem [{\citenamefont {Yang}\ and\ \citenamefont {Kim}(2010)}]{BJYang}%
  \BibitemOpen
  \bibfield  {author} {\bibinfo {author} {\bibfnamefont {Bohm-Jung}\
  \bibnamefont {Yang}}\ and\ \bibinfo {author} {\bibfnamefont {Yong~Baek}\
  \bibnamefont {Kim}},\ }\bibfield  {title} {\enquote {\bibinfo {title}
  {Topological insulators and metal-insulator transition in the pyrochlore
  iridates},}\ }\href {\doibase 10.1103/PhysRevB.82.085111} {\bibfield
  {journal} {\bibinfo  {journal} {Phys. Rev. B}\ }\textbf {\bibinfo {volume}
  {82}},\ \bibinfo {pages} {085111} (\bibinfo {year} {2010})}\BibitemShut
  {NoStop}%
\bibitem [{\citenamefont {Moon}\ \emph {et~al.}(2013)\citenamefont {Moon},
  \citenamefont {Xu}, \citenamefont {Kim},\ and\ \citenamefont
  {Balents}}]{EGMoon}%
  \BibitemOpen
  \bibfield  {author} {\bibinfo {author} {\bibfnamefont {Eun-Gook}\
  \bibnamefont {Moon}}, \bibinfo {author} {\bibfnamefont {Cenke}\ \bibnamefont
  {Xu}}, \bibinfo {author} {\bibfnamefont {Yong~Baek}\ \bibnamefont {Kim}}, \
  and\ \bibinfo {author} {\bibfnamefont {Leon}\ \bibnamefont {Balents}},\
  }\bibfield  {title} {\enquote {\bibinfo {title} {{Non-Fermi-Liquid and
  Topological States with Strong Spin-Orbit Coupling}},}\ }\href {\doibase
  10.1103/PhysRevLett.111.206401} {\bibfield  {journal} {\bibinfo  {journal}
  {Phys. Rev. Lett.}\ }\textbf {\bibinfo {volume} {111}},\ \bibinfo {pages}
  {206401} (\bibinfo {year} {2013})}\BibitemShut {NoStop}%
\bibitem [{\citenamefont {Witczak-Krempa}\ and\ \citenamefont
  {Kim}(2012)}]{YBKim1}%
  \BibitemOpen
  \bibfield  {author} {\bibinfo {author} {\bibfnamefont {William}\ \bibnamefont
  {Witczak-Krempa}}\ and\ \bibinfo {author} {\bibfnamefont {Yong~Baek}\
  \bibnamefont {Kim}},\ }\bibfield  {title} {\enquote {\bibinfo {title}
  {Topological and magnetic phases of interacting electrons in the pyrochlore
  iridates},}\ }\href {\doibase 10.1103/PhysRevB.85.045124} {\bibfield
  {journal} {\bibinfo  {journal} {Phys. Rev. B}\ }\textbf {\bibinfo {volume}
  {85}},\ \bibinfo {pages} {045124} (\bibinfo {year} {2012})}\BibitemShut
  {NoStop}%
\bibitem [{\citenamefont {Go}\ \emph {et~al.}(2012)\citenamefont {Go},
  \citenamefont {Witczak-Krempa}, \citenamefont {Jeon}, \citenamefont {Park},\
  and\ \citenamefont {Kim}}]{YBKim2}%
  \BibitemOpen
  \bibfield  {author} {\bibinfo {author} {\bibfnamefont {Ara}\ \bibnamefont
  {Go}}, \bibinfo {author} {\bibfnamefont {William}\ \bibnamefont
  {Witczak-Krempa}}, \bibinfo {author} {\bibfnamefont {Gun~Sang}\ \bibnamefont
  {Jeon}}, \bibinfo {author} {\bibfnamefont {Kwon}\ \bibnamefont {Park}}, \
  and\ \bibinfo {author} {\bibfnamefont {Yong~Baek}\ \bibnamefont {Kim}},\
  }\bibfield  {title} {\enquote {\bibinfo {title} {{Correlation Effects on 3D
  Topological Phases: From Bulk to Boundary}},}\ }\href {\doibase
  10.1103/PhysRevLett.109.066401} {\bibfield  {journal} {\bibinfo  {journal}
  {Phys. Rev. Lett.}\ }\textbf {\bibinfo {volume} {109}},\ \bibinfo {pages}
  {066401} (\bibinfo {year} {2012})}\BibitemShut {NoStop}%
\bibitem [{\citenamefont {Chen}\ and\ \citenamefont
  {Hermele}(2012)}]{ChenHermele}%
  \BibitemOpen
  \bibfield  {author} {\bibinfo {author} {\bibfnamefont {Gang}\ \bibnamefont
  {Chen}}\ and\ \bibinfo {author} {\bibfnamefont {Michael}\ \bibnamefont
  {Hermele}},\ }\bibfield  {title} {\enquote {\bibinfo {title} {Magnetic orders
  and topological phases from $f$-$d$ exchange in pyrochlore iridates},}\
  }\href {\doibase 10.1103/PhysRevB.86.235129} {\bibfield  {journal} {\bibinfo
  {journal} {Phys. Rev. B}\ }\textbf {\bibinfo {volume} {86}},\ \bibinfo
  {pages} {235129} (\bibinfo {year} {2012})}\BibitemShut {NoStop}%
\bibitem [{\citenamefont {Lee}\ \emph {et~al.}(2013)\citenamefont {Lee},
  \citenamefont {Paramekanti},\ and\ \citenamefont {Kim}}]{LeeArun}%
  \BibitemOpen
  \bibfield  {author} {\bibinfo {author} {\bibfnamefont {SungBin}\ \bibnamefont
  {Lee}}, \bibinfo {author} {\bibfnamefont {Arun}\ \bibnamefont {Paramekanti}},
  \ and\ \bibinfo {author} {\bibfnamefont {Yong~Baek}\ \bibnamefont {Kim}},\
  }\bibfield  {title} {\enquote {\bibinfo {title} {{RKKY Interactions and the
  Anomalous Hall Effect in Metallic Rare-Earth Pyrochlores}},}\ }\href
  {\doibase 10.1103/PhysRevLett.111.196601} {\bibfield  {journal} {\bibinfo
  {journal} {Phys. Rev. Lett.}\ }\textbf {\bibinfo {volume} {111}},\ \bibinfo
  {pages} {196601} (\bibinfo {year} {2013})}\BibitemShut {NoStop}%
\bibitem [{\citenamefont {Savary}\ \emph {et~al.}(2014)\citenamefont {Savary},
  \citenamefont {Moon},\ and\ \citenamefont {Balents}}]{SavaryMoon}%
  \BibitemOpen
  \bibfield  {author} {\bibinfo {author} {\bibfnamefont {Lucile}\ \bibnamefont
  {Savary}}, \bibinfo {author} {\bibfnamefont {Eun-Gook}\ \bibnamefont {Moon}},
  \ and\ \bibinfo {author} {\bibfnamefont {Leon}\ \bibnamefont {Balents}},\
  }\bibfield  {title} {\enquote {\bibinfo {title} {{New Type of Quantum
  Criticality in the Pyrochlore Iridates}},}\ }\href {\doibase
  10.1103/PhysRevX.4.041027} {\bibfield  {journal} {\bibinfo  {journal} {Phys.
  Rev. X}\ }\textbf {\bibinfo {volume} {4}},\ \bibinfo {pages} {041027}
  (\bibinfo {year} {2014})}\BibitemShut {NoStop}%
\bibitem [{\citenamefont {Witczak-Krempa}\ \emph {et~al.}(2014)\citenamefont
  {Witczak-Krempa}, \citenamefont {Chen}, \citenamefont {Kim},\ and\
  \citenamefont {Balents}}]{WCKB}%
  \BibitemOpen
  \bibfield  {author} {\bibinfo {author} {\bibfnamefont {William}\ \bibnamefont
  {Witczak-Krempa}}, \bibinfo {author} {\bibfnamefont {Gang}\ \bibnamefont
  {Chen}}, \bibinfo {author} {\bibfnamefont {Yong~Baek}\ \bibnamefont {Kim}}, \
  and\ \bibinfo {author} {\bibfnamefont {Leon}\ \bibnamefont {Balents}},\
  }\bibfield  {title} {\enquote {\bibinfo {title} {Correlated quantum phenomena
  in the strong spin-orbit regime},}\ }\href@noop {} {\bibfield  {journal}
  {\bibinfo  {journal} {Annual Review of Condensed Matter Physics}\ }\textbf
  {\bibinfo {volume} {5}},\ \bibinfo {pages} {57--82} (\bibinfo {year}
  {2014})}\BibitemShut {NoStop}%
\bibitem [{\citenamefont {Wang}\ \emph {et~al.}(2017)\citenamefont {Wang},
  \citenamefont {Go},\ and\ \citenamefont {Millis}}]{Millis}%
  \BibitemOpen
  \bibfield  {author} {\bibinfo {author} {\bibfnamefont {Runzhi}\ \bibnamefont
  {Wang}}, \bibinfo {author} {\bibfnamefont {Ara}\ \bibnamefont {Go}}, \ and\
  \bibinfo {author} {\bibfnamefont {Andrew~J.}\ \bibnamefont {Millis}},\
  }\bibfield  {title} {\enquote {\bibinfo {title} {Electron interactions,
  spin-orbit coupling, and intersite correlations in pyrochlore iridates},}\
  }\href {\doibase 10.1103/PhysRevB.95.045133} {\bibfield  {journal} {\bibinfo
  {journal} {Phys. Rev. B}\ }\textbf {\bibinfo {volume} {95}},\ \bibinfo
  {pages} {045133} (\bibinfo {year} {2017})}\BibitemShut {NoStop}%
\bibitem [{\citenamefont {Lambert}\ \emph {et~al.}(2016)\citenamefont
  {Lambert}, \citenamefont {Schnyder}, \citenamefont {Moessner},\ and\
  \citenamefont {Eremin}}]{MoessnerSchnyder}%
  \BibitemOpen
  \bibfield  {author} {\bibinfo {author} {\bibfnamefont {F.}~\bibnamefont
  {Lambert}}, \bibinfo {author} {\bibfnamefont {A.~P.}\ \bibnamefont
  {Schnyder}}, \bibinfo {author} {\bibfnamefont {R.}~\bibnamefont {Moessner}},
  \ and\ \bibinfo {author} {\bibfnamefont {I.}~\bibnamefont {Eremin}},\
  }\bibfield  {title} {\enquote {\bibinfo {title} {{Quasiparticle interference
  from different impurities on the surface of pyrochlore iridates: Signatures
  of the Weyl phase}},}\ }\href {\doibase 10.1103/PhysRevB.94.165146}
  {\bibfield  {journal} {\bibinfo  {journal} {Phys. Rev. B}\ }\textbf {\bibinfo
  {volume} {94}},\ \bibinfo {pages} {165146} (\bibinfo {year}
  {2016})}\BibitemShut {NoStop}%
\bibitem [{\citenamefont {Tafti}\ \emph {et~al.}(2012)\citenamefont {Tafti},
  \citenamefont {Ishikawa}, \citenamefont {McCollam}, \citenamefont
  {Nakatsuji},\ and\ \citenamefont {Julian}}]{PhysRevB.85.205104}%
  \BibitemOpen
  \bibfield  {author} {\bibinfo {author} {\bibfnamefont {F.~F.}\ \bibnamefont
  {Tafti}}, \bibinfo {author} {\bibfnamefont {J.~J.}\ \bibnamefont {Ishikawa}},
  \bibinfo {author} {\bibfnamefont {A.}~\bibnamefont {McCollam}}, \bibinfo
  {author} {\bibfnamefont {S.}~\bibnamefont {Nakatsuji}}, \ and\ \bibinfo
  {author} {\bibfnamefont {S.~R.}\ \bibnamefont {Julian}},\ }\bibfield  {title}
  {\enquote {\bibinfo {title} {{Pressure-tuned insulator to metal transition in
  ${\text{Eu}}_{{2}}{\text{Ir}}_{{2}}{\text{O}}_{{7}}$}},}\ }\href {\doibase
  10.1103/PhysRevB.85.205104} {\bibfield  {journal} {\bibinfo  {journal} {Phys.
  Rev. B}\ }\textbf {\bibinfo {volume} {85}},\ \bibinfo {pages} {205104}
  (\bibinfo {year} {2012})}\BibitemShut {NoStop}%
\bibitem [{\citenamefont {Ishikawa}\ \emph {et~al.}(2012)\citenamefont
  {Ishikawa}, \citenamefont {O'Farrell},\ and\ \citenamefont
  {Nakatsuji}}]{PhysRevB.85.245109}%
  \BibitemOpen
  \bibfield  {author} {\bibinfo {author} {\bibfnamefont {Jun~J.}\ \bibnamefont
  {Ishikawa}}, \bibinfo {author} {\bibfnamefont {Eoin C.~T.}\ \bibnamefont
  {O'Farrell}}, \ and\ \bibinfo {author} {\bibfnamefont {Satoru}\ \bibnamefont
  {Nakatsuji}},\ }\bibfield  {title} {\enquote {\bibinfo {title} {{Continuous
  transition between antiferromagnetic insulator and paramagnetic metal in the
  pyrochlore iridate Eu${}_{2}$Ir${}_{2}$O${}_{7}$}},}\ }\href {\doibase
  10.1103/PhysRevB.85.245109} {\bibfield  {journal} {\bibinfo  {journal} {Phys.
  Rev. B}\ }\textbf {\bibinfo {volume} {85}},\ \bibinfo {pages} {245109}
  (\bibinfo {year} {2012})}\BibitemShut {NoStop}%
\bibitem [{\citenamefont {Zhao}\ \emph {et~al.}(2011)\citenamefont {Zhao},
  \citenamefont {Mackie}, \citenamefont {MacLaughlin}, \citenamefont {Bernal},
  \citenamefont {Ishikawa}, \citenamefont {Ohta},\ and\ \citenamefont
  {Nakatsuji}}]{PhysRevB.83.180402}%
  \BibitemOpen
  \bibfield  {author} {\bibinfo {author} {\bibfnamefont {Songrui}\ \bibnamefont
  {Zhao}}, \bibinfo {author} {\bibfnamefont {J.~M.}\ \bibnamefont {Mackie}},
  \bibinfo {author} {\bibfnamefont {D.~E.}\ \bibnamefont {MacLaughlin}},
  \bibinfo {author} {\bibfnamefont {O.~O.}\ \bibnamefont {Bernal}}, \bibinfo
  {author} {\bibfnamefont {J.~J.}\ \bibnamefont {Ishikawa}}, \bibinfo {author}
  {\bibfnamefont {Y.}~\bibnamefont {Ohta}}, \ and\ \bibinfo {author}
  {\bibfnamefont {S.}~\bibnamefont {Nakatsuji}},\ }\bibfield  {title} {\enquote
  {\bibinfo {title} {{Magnetic transition, long-range order, and moment
  fluctuations in the pyrochlore iridate Eu${}_{2}$Ir${}_{2}$O${}_{7}$}},}\
  }\href {\doibase 10.1103/PhysRevB.83.180402} {\bibfield  {journal} {\bibinfo
  {journal} {Phys. Rev. B}\ }\textbf {\bibinfo {volume} {83}},\ \bibinfo
  {pages} {180402} (\bibinfo {year} {2011})}\BibitemShut {NoStop}%
\bibitem [{\citenamefont {Disseler}\ \emph
  {et~al.}(2012{\natexlab{a}})\citenamefont {Disseler}, \citenamefont {Dhital},
  \citenamefont {Amato}, \citenamefont {Giblin}, \citenamefont {de~la Cruz},
  \citenamefont {Wilson},\ and\ \citenamefont {Graf}}]{PhysRevB.86.014428}%
  \BibitemOpen
  \bibfield  {author} {\bibinfo {author} {\bibfnamefont {S.~M.}\ \bibnamefont
  {Disseler}}, \bibinfo {author} {\bibfnamefont {Chetan}\ \bibnamefont
  {Dhital}}, \bibinfo {author} {\bibfnamefont {A.}~\bibnamefont {Amato}},
  \bibinfo {author} {\bibfnamefont {S.~R.}\ \bibnamefont {Giblin}}, \bibinfo
  {author} {\bibfnamefont {Clarina}\ \bibnamefont {de~la Cruz}}, \bibinfo
  {author} {\bibfnamefont {Stephen~D.}\ \bibnamefont {Wilson}}, \ and\ \bibinfo
  {author} {\bibfnamefont {M.~J.}\ \bibnamefont {Graf}},\ }\bibfield  {title}
  {\enquote {\bibinfo {title} {{Magnetic order in the pyrochlore iridates
  ${A}_{2}$Ir${}_{2}$O${}_{7}$ ($A$ = Y, Yb)}},}\ }\href {\doibase
  10.1103/PhysRevB.86.014428} {\bibfield  {journal} {\bibinfo  {journal} {Phys.
  Rev. B}\ }\textbf {\bibinfo {volume} {86}},\ \bibinfo {pages} {014428}
  (\bibinfo {year} {2012}{\natexlab{a}})}\BibitemShut {NoStop}%
\bibitem [{\citenamefont {Disseler}\ \emph
  {et~al.}(2012{\natexlab{b}})\citenamefont {Disseler}, \citenamefont {Dhital},
  \citenamefont {Hogan}, \citenamefont {Amato}, \citenamefont {Giblin},
  \citenamefont {de~la Cruz}, \citenamefont {Daoud-Aladine}, \citenamefont
  {Wilson},\ and\ \citenamefont {Graf}}]{PhysRevB.85.174441}%
  \BibitemOpen
  \bibfield  {author} {\bibinfo {author} {\bibfnamefont {S.~M.}\ \bibnamefont
  {Disseler}}, \bibinfo {author} {\bibfnamefont {Chetan}\ \bibnamefont
  {Dhital}}, \bibinfo {author} {\bibfnamefont {T.~C.}\ \bibnamefont {Hogan}},
  \bibinfo {author} {\bibfnamefont {A.}~\bibnamefont {Amato}}, \bibinfo
  {author} {\bibfnamefont {S.~R.}\ \bibnamefont {Giblin}}, \bibinfo {author}
  {\bibfnamefont {Clarina}\ \bibnamefont {de~la Cruz}}, \bibinfo {author}
  {\bibfnamefont {A.}~\bibnamefont {Daoud-Aladine}}, \bibinfo {author}
  {\bibfnamefont {Stephen~D.}\ \bibnamefont {Wilson}}, \ and\ \bibinfo {author}
  {\bibfnamefont {M.~J.}\ \bibnamefont {Graf}},\ }\bibfield  {title} {\enquote
  {\bibinfo {title} {{Magnetic order and the electronic ground state in the
  pyrochlore iridate Nd${}_{2}$Ir${}_{2}$O${}_{7}$}},}\ }\href {\doibase
  10.1103/PhysRevB.85.174441} {\bibfield  {journal} {\bibinfo  {journal} {Phys.
  Rev. B}\ }\textbf {\bibinfo {volume} {85}},\ \bibinfo {pages} {174441}
  (\bibinfo {year} {2012}{\natexlab{b}})}\BibitemShut {NoStop}%
\bibitem [{\citenamefont {Zhu}\ \emph {et~al.}(2014)\citenamefont {Zhu},
  \citenamefont {Wang}, \citenamefont {Seradjeh}, \citenamefont {Yang},\ and\
  \citenamefont {Zhang}}]{PhysRevB.90.054419}%
  \BibitemOpen
  \bibfield  {author} {\bibinfo {author} {\bibfnamefont {W.~K.}\ \bibnamefont
  {Zhu}}, \bibinfo {author} {\bibfnamefont {M.}~\bibnamefont {Wang}}, \bibinfo
  {author} {\bibfnamefont {B.}~\bibnamefont {Seradjeh}}, \bibinfo {author}
  {\bibfnamefont {Fengyuan}\ \bibnamefont {Yang}}, \ and\ \bibinfo {author}
  {\bibfnamefont {S.~X.}\ \bibnamefont {Zhang}},\ }\bibfield  {title} {\enquote
  {\bibinfo {title} {{Enhanced weak ferromagnetism and conductivity in
  hole-doped pyrochlore iridate
  ${\mathrm{Y}}_{2}{\mathrm{Ir}}_{2}{\mathrm{O}}_{7}$}},}\ }\href {\doibase
  10.1103/PhysRevB.90.054419} {\bibfield  {journal} {\bibinfo  {journal} {Phys.
  Rev. B}\ }\textbf {\bibinfo {volume} {90}},\ \bibinfo {pages} {054419}
  (\bibinfo {year} {2014})}\BibitemShut {NoStop}%
\bibitem [{\citenamefont {Takatsu}\ \emph {et~al.}(2014)\citenamefont
  {Takatsu}, \citenamefont {Watanabe}, \citenamefont {Goto},\ and\
  \citenamefont {Kadowaki}}]{PhysRevB.90.235110}%
  \BibitemOpen
  \bibfield  {author} {\bibinfo {author} {\bibfnamefont {Hiroshi}\ \bibnamefont
  {Takatsu}}, \bibinfo {author} {\bibfnamefont {Kunihiko}\ \bibnamefont
  {Watanabe}}, \bibinfo {author} {\bibfnamefont {Kazuki}\ \bibnamefont {Goto}},
  \ and\ \bibinfo {author} {\bibfnamefont {Hiroaki}\ \bibnamefont {Kadowaki}},\
  }\bibfield  {title} {\enquote {\bibinfo {title} {{Comparative study of
  low-temperature x-ray diffraction experiments on
  ${R}_{2}{\mathrm{Ir}}_{2}{\mathrm{O}}_{7}$ ($R=\mathrm{Nd}$, Eu, and Pr)}},}\
  }\href {\doibase 10.1103/PhysRevB.90.235110} {\bibfield  {journal} {\bibinfo
  {journal} {Phys. Rev. B}\ }\textbf {\bibinfo {volume} {90}},\ \bibinfo
  {pages} {235110} (\bibinfo {year} {2014})}\BibitemShut {NoStop}%
\bibitem [{\citenamefont {Prando}\ \emph {et~al.}(2016)\citenamefont {Prando},
  \citenamefont {Dally}, \citenamefont {Schottenhamel}, \citenamefont
  {Guguchia}, \citenamefont {Baek}, \citenamefont {Aeschlimann}, \citenamefont
  {Wolter}, \citenamefont {Wilson}, \citenamefont {B\"uchner},\ and\
  \citenamefont {Graf}}]{PhysRevB.93.104422}%
  \BibitemOpen
  \bibfield  {author} {\bibinfo {author} {\bibfnamefont {G.}~\bibnamefont
  {Prando}}, \bibinfo {author} {\bibfnamefont {R.}~\bibnamefont {Dally}},
  \bibinfo {author} {\bibfnamefont {W.}~\bibnamefont {Schottenhamel}}, \bibinfo
  {author} {\bibfnamefont {Z.}~\bibnamefont {Guguchia}}, \bibinfo {author}
  {\bibfnamefont {S.-H.}\ \bibnamefont {Baek}}, \bibinfo {author}
  {\bibfnamefont {R.}~\bibnamefont {Aeschlimann}}, \bibinfo {author}
  {\bibfnamefont {A.~U.~B.}\ \bibnamefont {Wolter}}, \bibinfo {author}
  {\bibfnamefont {S.~D.}\ \bibnamefont {Wilson}}, \bibinfo {author}
  {\bibfnamefont {B.}~\bibnamefont {B\"uchner}}, \ and\ \bibinfo {author}
  {\bibfnamefont {M.~J.}\ \bibnamefont {Graf}},\ }\bibfield  {title} {\enquote
  {\bibinfo {title} {{Influence of hydrostatic pressure on the bulk magnetic
  properties of ${\mathrm{Eu}}_{2}{\mathrm{Ir}}_{2}{\mathrm{O}}_{7}$}},}\
  }\href {\doibase 10.1103/PhysRevB.93.104422} {\bibfield  {journal} {\bibinfo
  {journal} {Phys. Rev. B}\ }\textbf {\bibinfo {volume} {93}},\ \bibinfo
  {pages} {104422} (\bibinfo {year} {2016})}\BibitemShut {NoStop}%
\bibitem [{\citenamefont {Flint}\ and\ \citenamefont
  {Senthil}(2013)}]{PhysRevB.87.125147}%
  \BibitemOpen
  \bibfield  {author} {\bibinfo {author} {\bibfnamefont {Rebecca}\ \bibnamefont
  {Flint}}\ and\ \bibinfo {author} {\bibfnamefont {T.}~\bibnamefont
  {Senthil}},\ }\bibfield  {title} {\enquote {\bibinfo {title} {{Chiral RKKY
  interaction in Pr${}_{2}$Ir${}_{2}$O${}_{7}$}},}\ }\href {\doibase
  10.1103/PhysRevB.87.125147} {\bibfield  {journal} {\bibinfo  {journal} {Phys.
  Rev. B}\ }\textbf {\bibinfo {volume} {87}},\ \bibinfo {pages} {125147}
  (\bibinfo {year} {2013})}\BibitemShut {NoStop}%
\bibitem [{\citenamefont {Tian}\ \emph {et~al.}(2016)\citenamefont {Tian},
  \citenamefont {Kohama}, \citenamefont {Tomita}, \citenamefont {Ishizuka},
  \citenamefont {Hsieh}, \citenamefont {Ishikawa}, \citenamefont {Kindo},
  \citenamefont {Balents},\ and\ \citenamefont {Nakatsuji}}]{Nd2Ir2O7}%
  \BibitemOpen
  \bibfield  {author} {\bibinfo {author} {\bibfnamefont {Zhaoming}\
  \bibnamefont {Tian}}, \bibinfo {author} {\bibfnamefont {Yoshimitsu}\
  \bibnamefont {Kohama}}, \bibinfo {author} {\bibfnamefont {Takahiro}\
  \bibnamefont {Tomita}}, \bibinfo {author} {\bibfnamefont {Hiroaki}\
  \bibnamefont {Ishizuka}}, \bibinfo {author} {\bibfnamefont {Timothy~H.}\
  \bibnamefont {Hsieh}}, \bibinfo {author} {\bibfnamefont {Jun~J.}\
  \bibnamefont {Ishikawa}}, \bibinfo {author} {\bibfnamefont {Koichi}\
  \bibnamefont {Kindo}}, \bibinfo {author} {\bibfnamefont {Leon}\ \bibnamefont
  {Balents}}, \ and\ \bibinfo {author} {\bibfnamefont {Satoru}\ \bibnamefont
  {Nakatsuji}},\ }\bibfield  {title} {\enquote {\bibinfo {title}
  {{Field-induced quantum metal–insulator transition in the pyrochlore
  iridate Nd$_2$Ir$_2$O$_7$}},}\ }\href {\doibase 10.1038/nphys3567} {\bibfield
   {journal} {\bibinfo  {journal} {Nature Physics}\ }\textbf {\bibinfo {volume}
  {12}},\ \bibinfo {pages} {134–138} (\bibinfo {year} {2016})}\BibitemShut
  {NoStop}%
\bibitem [{\citenamefont {Machida}\ \emph {et~al.}(2010)\citenamefont
  {Machida}, \citenamefont {Nakatsuji}, \citenamefont {Onoda}, \citenamefont
  {Tayama},\ and\ \citenamefont {Sakakibara}}]{Machida}%
  \BibitemOpen
  \bibfield  {author} {\bibinfo {author} {\bibfnamefont {Y.}~\bibnamefont
  {Machida}}, \bibinfo {author} {\bibfnamefont {S.}~\bibnamefont {Nakatsuji}},
  \bibinfo {author} {\bibfnamefont {S.}~\bibnamefont {Onoda}}, \bibinfo
  {author} {\bibfnamefont {T.}~\bibnamefont {Tayama}}, \ and\ \bibinfo {author}
  {\bibfnamefont {T}~\bibnamefont {Sakakibara}},\ }\bibfield  {title} {\enquote
  {\bibinfo {title} {Time-reversal symmetry breaking and spontaneous hall
  effect without magnetic dipole order},}\ }\href {\doibase
  10.1038/nature08680} {\bibfield  {journal} {\bibinfo  {journal} {Nature}\
  }\textbf {\bibinfo {volume} {463}},\ \bibinfo {pages} {210--213} (\bibinfo
  {year} {2010})}\BibitemShut {NoStop}%
\bibitem [{\citenamefont {Nakatsuji}\ \emph {et~al.}(2006)\citenamefont
  {Nakatsuji}, \citenamefont {Machida}, \citenamefont {Maeno}, \citenamefont
  {Tayama}, \citenamefont {Sakakibara}, \citenamefont {Duijn}, \citenamefont
  {Balicas}, \citenamefont {Millican}, \citenamefont {Macaluso},\ and\
  \citenamefont {Chan}}]{PhysRevLett.96.087204}%
  \BibitemOpen
  \bibfield  {author} {\bibinfo {author} {\bibfnamefont {S.}~\bibnamefont
  {Nakatsuji}}, \bibinfo {author} {\bibfnamefont {Y.}~\bibnamefont {Machida}},
  \bibinfo {author} {\bibfnamefont {Y.}~\bibnamefont {Maeno}}, \bibinfo
  {author} {\bibfnamefont {T.}~\bibnamefont {Tayama}}, \bibinfo {author}
  {\bibfnamefont {T.}~\bibnamefont {Sakakibara}}, \bibinfo {author}
  {\bibfnamefont {J.~van}\ \bibnamefont {Duijn}}, \bibinfo {author}
  {\bibfnamefont {L.}~\bibnamefont {Balicas}}, \bibinfo {author} {\bibfnamefont
  {J.~N.}\ \bibnamefont {Millican}}, \bibinfo {author} {\bibfnamefont {R.~T.}\
  \bibnamefont {Macaluso}}, \ and\ \bibinfo {author} {\bibfnamefont {Julia~Y.}\
  \bibnamefont {Chan}},\ }\bibfield  {title} {\enquote {\bibinfo {title}
  {{Metallic Spin-Liquid Behavior of the Geometrically Frustrated Kondo Lattice
  ${\mathrm{Pr}}_{2}{\mathrm{Ir}}_{2}{\mathrm{O}}_{7}$}},}\ }\href {\doibase
  10.1103/PhysRevLett.96.087204} {\bibfield  {journal} {\bibinfo  {journal}
  {Phys. Rev. Lett.}\ }\textbf {\bibinfo {volume} {96}},\ \bibinfo {pages}
  {087204} (\bibinfo {year} {2006})}\BibitemShut {NoStop}%
\bibitem [{\citenamefont {Tokiwa}\ \emph {et~al.}(2014)\citenamefont {Tokiwa},
  \citenamefont {Ishikawa}, \citenamefont {Nakatsuji},\ and\ \citenamefont
  {Gegenwart}}]{Gegenwart}%
  \BibitemOpen
  \bibfield  {author} {\bibinfo {author} {\bibfnamefont {Y.}~\bibnamefont
  {Tokiwa}}, \bibinfo {author} {\bibfnamefont {J.J.}\ \bibnamefont {Ishikawa}},
  \bibinfo {author} {\bibfnamefont {S.}~\bibnamefont {Nakatsuji}}, \ and\
  \bibinfo {author} {\bibfnamefont {P.}~\bibnamefont {Gegenwart}},\ }\bibfield
  {title} {\enquote {\bibinfo {title} {Quantum criticality in a metallic spin
  liquid},}\ }\href {\doibase 10.1038/nmat3900} {\bibfield  {journal} {\bibinfo
   {journal} {Nature Materials}\ }\textbf {\bibinfo {volume} {13}},\ \bibinfo
  {pages} {356–359} (\bibinfo {year} {2014})}\BibitemShut {NoStop}%
\bibitem [{\citenamefont {MacLaughlin}\ \emph {et~al.}(2015)\citenamefont
  {MacLaughlin}, \citenamefont {Bernal}, \citenamefont {Shu}, \citenamefont
  {Ishikawa}, \citenamefont {Matsumoto}, \citenamefont {Wen}, \citenamefont
  {Mourigal}, \citenamefont {Stock}, \citenamefont {Ehlers}, \citenamefont
  {Broholm}, \citenamefont {Machida}, \citenamefont {Kimura}, \citenamefont
  {Nakatsuji}, \citenamefont {Shimura},\ and\ \citenamefont
  {Sakakibara}}]{PhysRevB.92.054432}%
  \BibitemOpen
  \bibfield  {author} {\bibinfo {author} {\bibfnamefont {D.~E.}\ \bibnamefont
  {MacLaughlin}}, \bibinfo {author} {\bibfnamefont {O.~O.}\ \bibnamefont
  {Bernal}}, \bibinfo {author} {\bibfnamefont {Lei}\ \bibnamefont {Shu}},
  \bibinfo {author} {\bibfnamefont {Jun}\ \bibnamefont {Ishikawa}}, \bibinfo
  {author} {\bibfnamefont {Yosuke}\ \bibnamefont {Matsumoto}}, \bibinfo
  {author} {\bibfnamefont {J.-J.}\ \bibnamefont {Wen}}, \bibinfo {author}
  {\bibfnamefont {M.}~\bibnamefont {Mourigal}}, \bibinfo {author}
  {\bibfnamefont {C.}~\bibnamefont {Stock}}, \bibinfo {author} {\bibfnamefont
  {G.}~\bibnamefont {Ehlers}}, \bibinfo {author} {\bibfnamefont {C.~L.}\
  \bibnamefont {Broholm}}, \bibinfo {author} {\bibfnamefont {Yo}~\bibnamefont
  {Machida}}, \bibinfo {author} {\bibfnamefont {Kenta}\ \bibnamefont {Kimura}},
  \bibinfo {author} {\bibfnamefont {Satoru}\ \bibnamefont {Nakatsuji}},
  \bibinfo {author} {\bibfnamefont {Yasuyuki}\ \bibnamefont {Shimura}}, \ and\
  \bibinfo {author} {\bibfnamefont {Toshiro}\ \bibnamefont {Sakakibara}},\
  }\bibfield  {title} {\enquote {\bibinfo {title} {{Unstable spin-ice order in
  the stuffed metallic pyrochlore
  ${\mathrm{Pr}}_{2+x}{\mathrm{Ir}}_{2\ensuremath{-}x}{\mathrm{O}}_{7\ensuremath{-}\ensuremath{\delta}}$}},}\
  }\href {\doibase 10.1103/PhysRevB.92.054432} {\bibfield  {journal} {\bibinfo
  {journal} {Phys. Rev. B}\ }\textbf {\bibinfo {volume} {92}},\ \bibinfo
  {pages} {054432} (\bibinfo {year} {2015})}\BibitemShut {NoStop}%
\bibitem [{\citenamefont {Melko}\ \emph {et~al.}(2001)\citenamefont {Melko},
  \citenamefont {den Hertog},\ and\ \citenamefont
  {Gingras}}]{MelkoHertogGingras}%
  \BibitemOpen
  \bibfield  {author} {\bibinfo {author} {\bibfnamefont {Roger~G.}\
  \bibnamefont {Melko}}, \bibinfo {author} {\bibfnamefont {Byron~C.}\
  \bibnamefont {den Hertog}}, \ and\ \bibinfo {author} {\bibfnamefont {Michel
  J.~P.}\ \bibnamefont {Gingras}},\ }\bibfield  {title} {\enquote {\bibinfo
  {title} {{Long-Range Order at Low Temperatures in Dipolar Spin Ice}},}\
  }\href {\doibase 10.1103/PhysRevLett.87.067203} {\bibfield  {journal}
  {\bibinfo  {journal} {Phys. Rev. Lett.}\ }\textbf {\bibinfo {volume} {87}},\
  \bibinfo {pages} {067203} (\bibinfo {year} {2001})}\BibitemShut {NoStop}%
\bibitem [{\citenamefont {Chen}(2016)}]{GangChen2016}%
  \BibitemOpen
  \bibfield  {author} {\bibinfo {author} {\bibfnamefont {Gang}\ \bibnamefont
  {Chen}},\ }\bibfield  {title} {\enquote {\bibinfo {title} {{``Magnetic
  monopole'' condensation of the pyrochlore ice U(1) quantum spin liquid:
  Application to ${\mathrm{Pr}}_{2}{\mathrm{Ir}}_{2}{\mathrm{O}}_{7}$ and
  ${\mathrm{Yb}}_{2}{\mathrm{Ti}}_{2}{\mathrm{O}}_{7}$}},}\ }\href {\doibase
  10.1103/PhysRevB.94.205107} {\bibfield  {journal} {\bibinfo  {journal} {Phys.
  Rev. B}\ }\textbf {\bibinfo {volume} {94}},\ \bibinfo {pages} {205107}
  (\bibinfo {year} {2016})}\BibitemShut {NoStop}%
\bibitem [{\citenamefont {Kondo}\ \emph {et~al.}(2015)\citenamefont {Kondo},
  \citenamefont {Nakayama}, \citenamefont {Chen}, \citenamefont {Ishikawa},
  \citenamefont {Moon}, \citenamefont {Yamamoto}, \citenamefont {Ota},
  \citenamefont {Malaeb}, \citenamefont {Kanai}, \citenamefont {Nakashima},
  \citenamefont {Ishida}, \citenamefont {Yoshida}, \citenamefont {Yamamoto},
  \citenamefont {Matsunami}, \citenamefont {Kimura}, \citenamefont {Inami},
  \citenamefont {Ono}, \citenamefont {Kumigashira}, \citenamefont {Nakatsuji},
  \citenamefont {Balents},\ and\ \citenamefont {Shin}}]{Kondo2015}%
  \BibitemOpen
  \bibfield  {author} {\bibinfo {author} {\bibfnamefont {T}~\bibnamefont
  {Kondo}}, \bibinfo {author} {\bibfnamefont {M.}~\bibnamefont {Nakayama}},
  \bibinfo {author} {\bibfnamefont {R.}~\bibnamefont {Chen}}, \bibinfo {author}
  {\bibfnamefont {J.J.}\ \bibnamefont {Ishikawa}}, \bibinfo {author}
  {\bibfnamefont {E.-G.}\ \bibnamefont {Moon}}, \bibinfo {author}
  {\bibfnamefont {T.}~\bibnamefont {Yamamoto}}, \bibinfo {author}
  {\bibfnamefont {Y.}~\bibnamefont {Ota}}, \bibinfo {author} {\bibfnamefont
  {W.}~\bibnamefont {Malaeb}}, \bibinfo {author} {\bibfnamefont
  {H.}~\bibnamefont {Kanai}}, \bibinfo {author} {\bibfnamefont
  {Y.}~\bibnamefont {Nakashima}}, \bibinfo {author} {\bibfnamefont
  {Y.}~\bibnamefont {Ishida}}, \bibinfo {author} {\bibfnamefont
  {R.}~\bibnamefont {Yoshida}}, \bibinfo {author} {\bibfnamefont
  {H.}~\bibnamefont {Yamamoto}}, \bibinfo {author} {\bibfnamefont
  {M.}~\bibnamefont {Matsunami}}, \bibinfo {author} {\bibfnamefont
  {S.}~\bibnamefont {Kimura}}, \bibinfo {author} {\bibfnamefont
  {N.}~\bibnamefont {Inami}}, \bibinfo {author} {\bibfnamefont
  {K.}~\bibnamefont {Ono}}, \bibinfo {author} {\bibfnamefont {H.}~\bibnamefont
  {Kumigashira}}, \bibinfo {author} {\bibfnamefont {S.}~\bibnamefont
  {Nakatsuji}}, \bibinfo {author} {\bibfnamefont {L.}~\bibnamefont {Balents}},
  \ and\ \bibinfo {author} {\bibfnamefont {S.}~\bibnamefont {Shin}},\
  }\bibfield  {title} {\enquote {\bibinfo {title} {{Quadratic Fermi node in a
  3D strongly correlated semimetal}},}\ }\href {\doibase 10.1038/ncomms10042}
  {\bibfield  {journal} {\bibinfo  {journal} {Nature Communications}\ }\textbf
  {\bibinfo {volume} {6}},\ \bibinfo {pages} {10042} (\bibinfo {year}
  {2015})}\BibitemShut {NoStop}%
\bibitem [{\citenamefont {Ishii}\ \emph {et~al.}(2015)\citenamefont {Ishii},
  \citenamefont {Mizuta}, \citenamefont {Kato}, \citenamefont {Ozaki},
  \citenamefont {Weng},\ and\ \citenamefont {Onoda}}]{Onoda2015}%
  \BibitemOpen
  \bibfield  {author} {\bibinfo {author} {\bibfnamefont {Fumiyuki}\
  \bibnamefont {Ishii}}, \bibinfo {author} {\bibfnamefont {Yo~Pierre}\
  \bibnamefont {Mizuta}}, \bibinfo {author} {\bibfnamefont {Takehiro}\
  \bibnamefont {Kato}}, \bibinfo {author} {\bibfnamefont {Taisuke}\
  \bibnamefont {Ozaki}}, \bibinfo {author} {\bibfnamefont {Hongming}\
  \bibnamefont {Weng}}, \ and\ \bibinfo {author} {\bibfnamefont {Shigeki}\
  \bibnamefont {Onoda}},\ }\bibfield  {title} {\enquote {\bibinfo {title}
  {{First-Principles Study on Cubic Pyrochlore Iridates Y$_2$Ir$_2$O$_7$ and
  Pr$_2$Ir$_2$O$_7$}},}\ }\href {\doibase 10.7566/JPSJ.84.073703} {\bibfield
  {journal} {\bibinfo  {journal} {Journal of the Physical Society of Japan}\
  }\textbf {\bibinfo {volume} {84}},\ \bibinfo {pages} {073703} (\bibinfo
  {year} {2015})}\BibitemShut {NoStop}%
\bibitem [{\citenamefont {Cheng}\ \emph {et~al.}(2017)\citenamefont {Cheng},
  \citenamefont {Ohtsuki}, \citenamefont {Chaudhuri}, \citenamefont {Lippmaa},\
  and\ \citenamefont {Armitage}}]{Armitage2017}%
  \BibitemOpen
  \bibfield  {author} {\bibinfo {author} {\bibfnamefont {Bing}\ \bibnamefont
  {Cheng}}, \bibinfo {author} {\bibfnamefont {T.}~\bibnamefont {Ohtsuki}},
  \bibinfo {author} {\bibfnamefont {Dipanjan}\ \bibnamefont {Chaudhuri}},
  \bibinfo {author} {\bibfnamefont {Mikk}\ \bibnamefont {Lippmaa}}, \ and\
  \bibinfo {author} {\bibfnamefont {N.~P.}\ \bibnamefont {Armitage}},\
  }\bibfield  {title} {\enquote {\bibinfo {title} {Dielectric anomalies and
  interactions in the three-dimensional quadratic band touching luttinger
  semimetal pr$_2$ir$_2$o$_7$},}\ }\href {\doibase 10.1038/s41467-017-02121-y}
  {\bibfield  {journal} {\bibinfo  {journal} {Nature Communications}\ }\textbf
  {\bibinfo {volume} {8}},\ \bibinfo {pages} {2097} (\bibinfo {year}
  {2017})}\BibitemShut {NoStop}%
\bibitem [{\citenamefont {Bernevig}\ \emph {et~al.}(2006)\citenamefont
  {Bernevig}, \citenamefont {Hughes},\ and\ \citenamefont
  {Zhang}}]{Bernevig1757}%
  \BibitemOpen
  \bibfield  {author} {\bibinfo {author} {\bibfnamefont {B.~Andrei}\
  \bibnamefont {Bernevig}}, \bibinfo {author} {\bibfnamefont {Taylor~L.}\
  \bibnamefont {Hughes}}, \ and\ \bibinfo {author} {\bibfnamefont {Shou-Cheng}\
  \bibnamefont {Zhang}},\ }\bibfield  {title} {\enquote {\bibinfo {title}
  {{Quantum Spin Hall Effect and Topological Phase Transition in HgTe Quantum
  Wells}},}\ }\href {\doibase 10.1126/science.1133734} {\bibfield  {journal}
  {\bibinfo  {journal} {Science}\ }\textbf {\bibinfo {volume} {314}},\ \bibinfo
  {pages} {1757--1761} (\bibinfo {year} {2006})}\BibitemShut {NoStop}%
\bibitem [{\citenamefont {Br\"une}\ \emph {et~al.}(2011)\citenamefont
  {Br\"une}, \citenamefont {Liu}, \citenamefont {Novik}, \citenamefont
  {Hankiewicz}, \citenamefont {Buhmann}, \citenamefont {Chen}, \citenamefont
  {Qi}, \citenamefont {Shen}, \citenamefont {Zhang},\ and\ \citenamefont
  {Molenkamp}}]{PhysRevLett.106.126803}%
  \BibitemOpen
  \bibfield  {author} {\bibinfo {author} {\bibfnamefont {C.}~\bibnamefont
  {Br\"une}}, \bibinfo {author} {\bibfnamefont {C.~X.}\ \bibnamefont {Liu}},
  \bibinfo {author} {\bibfnamefont {E.~G.}\ \bibnamefont {Novik}}, \bibinfo
  {author} {\bibfnamefont {E.~M.}\ \bibnamefont {Hankiewicz}}, \bibinfo
  {author} {\bibfnamefont {H.}~\bibnamefont {Buhmann}}, \bibinfo {author}
  {\bibfnamefont {Y.~L.}\ \bibnamefont {Chen}}, \bibinfo {author}
  {\bibfnamefont {X.~L.}\ \bibnamefont {Qi}}, \bibinfo {author} {\bibfnamefont
  {Z.~X.}\ \bibnamefont {Shen}}, \bibinfo {author} {\bibfnamefont {S.~C.}\
  \bibnamefont {Zhang}}, \ and\ \bibinfo {author} {\bibfnamefont {L.~W.}\
  \bibnamefont {Molenkamp}},\ }\bibfield  {title} {\enquote {\bibinfo {title}
  {{Quantum Hall Effect from the Topological Surface States of Strained Bulk
  HgTe}},}\ }\href {\doibase 10.1103/PhysRevLett.106.126803} {\bibfield
  {journal} {\bibinfo  {journal} {Phys. Rev. Lett.}\ }\textbf {\bibinfo
  {volume} {106}},\ \bibinfo {pages} {126803} (\bibinfo {year}
  {2011})}\BibitemShut {NoStop}%
\bibitem [{\citenamefont {Kim}\ \emph {et~al.}(2008)\citenamefont {Kim},
  \citenamefont {Jin}, \citenamefont {Moon}, \citenamefont {Kim}, \citenamefont
  {Park}, \citenamefont {Leem}, \citenamefont {Yu}, \citenamefont {Noh},
  \citenamefont {Kim}, \citenamefont {Oh}, \citenamefont {Park}, \citenamefont
  {Durairaj}, \citenamefont {Cao},\ and\ \citenamefont {Rotenberg}}]{BJKim}%
  \BibitemOpen
  \bibfield  {author} {\bibinfo {author} {\bibfnamefont {B.~J.}\ \bibnamefont
  {Kim}}, \bibinfo {author} {\bibfnamefont {Hosub}\ \bibnamefont {Jin}},
  \bibinfo {author} {\bibfnamefont {S.~J.}\ \bibnamefont {Moon}}, \bibinfo
  {author} {\bibfnamefont {J.-Y.}\ \bibnamefont {Kim}}, \bibinfo {author}
  {\bibfnamefont {B.-G.}\ \bibnamefont {Park}}, \bibinfo {author}
  {\bibfnamefont {C.~S.}\ \bibnamefont {Leem}}, \bibinfo {author}
  {\bibfnamefont {Jaejun}\ \bibnamefont {Yu}}, \bibinfo {author} {\bibfnamefont
  {T.~W.}\ \bibnamefont {Noh}}, \bibinfo {author} {\bibfnamefont
  {C.}~\bibnamefont {Kim}}, \bibinfo {author} {\bibfnamefont {S.-J.}\
  \bibnamefont {Oh}}, \bibinfo {author} {\bibfnamefont {J.-H.}\ \bibnamefont
  {Park}}, \bibinfo {author} {\bibfnamefont {V.}~\bibnamefont {Durairaj}},
  \bibinfo {author} {\bibfnamefont {G.}~\bibnamefont {Cao}}, \ and\ \bibinfo
  {author} {\bibfnamefont {E.}~\bibnamefont {Rotenberg}},\ }\bibfield  {title}
  {\enquote {\bibinfo {title} {{Novel ${J}_{\mathrm{eff}}=1/2$ Mott State
  Induced by Relativistic Spin-Orbit Coupling in
  ${\mathrm{Sr}}_{2}{\mathrm{IrO}}_{4}$}},}\ }\href {\doibase
  10.1103/PhysRevLett.101.076402} {\bibfield  {journal} {\bibinfo  {journal}
  {Phys. Rev. Lett.}\ }\textbf {\bibinfo {volume} {101}},\ \bibinfo {pages}
  {076402} (\bibinfo {year} {2008})}\BibitemShut {NoStop}%
\bibitem [{\citenamefont {Chen}\ and\ \citenamefont
  {Balents}(2008)}]{PhysRevB.78.094403}%
  \BibitemOpen
  \bibfield  {author} {\bibinfo {author} {\bibfnamefont {Gang}\ \bibnamefont
  {Chen}}\ and\ \bibinfo {author} {\bibfnamefont {Leon}\ \bibnamefont
  {Balents}},\ }\bibfield  {title} {\enquote {\bibinfo {title} {{Spin-orbit
  effects in ${\text{Na}}_{4}{\text{Ir}}_{3}{\text{O}}_{8}$: A hyper-kagome
  lattice antiferromagnet}},}\ }\href {\doibase 10.1103/PhysRevB.78.094403}
  {\bibfield  {journal} {\bibinfo  {journal} {Phys. Rev. B}\ }\textbf {\bibinfo
  {volume} {78}},\ \bibinfo {pages} {094403} (\bibinfo {year}
  {2008})}\BibitemShut {NoStop}%
\bibitem [{\citenamefont {Chaloupka}\ \emph {et~al.}(2010)\citenamefont
  {Chaloupka}, \citenamefont {Jackeli},\ and\ \citenamefont
  {Khaliullin}}]{Khaliullin}%
  \BibitemOpen
  \bibfield  {author} {\bibinfo {author} {\bibfnamefont {Ji\ifmmode
  \check{r}\else~\v{r}\fi{}\'{\i}}\ \bibnamefont {Chaloupka}}, \bibinfo
  {author} {\bibfnamefont {George}\ \bibnamefont {Jackeli}}, \ and\ \bibinfo
  {author} {\bibfnamefont {Giniyat}\ \bibnamefont {Khaliullin}},\ }\bibfield
  {title} {\enquote {\bibinfo {title} {{Kitaev-Heisenberg Model on a Honeycomb
  Lattice: Possible Exotic Phases in Iridium Oxides
  ${A}_{2}{\mathrm{IrO}}_{3}$}},}\ }\href {\doibase
  10.1103/PhysRevLett.105.027204} {\bibfield  {journal} {\bibinfo  {journal}
  {Phys. Rev. Lett.}\ }\textbf {\bibinfo {volume} {105}},\ \bibinfo {pages}
  {027204} (\bibinfo {year} {2010})}\BibitemShut {NoStop}%
\bibitem [{\citenamefont {Goswami}\ \emph {et~al.}(2017)\citenamefont
  {Goswami}, \citenamefont {Roy},\ and\ \citenamefont
  {Das~Sarma}}]{PhysRevB.95.085120}%
  \BibitemOpen
  \bibfield  {author} {\bibinfo {author} {\bibfnamefont {Pallab}\ \bibnamefont
  {Goswami}}, \bibinfo {author} {\bibfnamefont {Bitan}\ \bibnamefont {Roy}}, \
  and\ \bibinfo {author} {\bibfnamefont {Sankar}\ \bibnamefont {Das~Sarma}},\
  }\bibfield  {title} {\enquote {\bibinfo {title} {Competing orders and
  topology in the global phase diagram of pyrochlore iridates},}\ }\href
  {\doibase 10.1103/PhysRevB.95.085120} {\bibfield  {journal} {\bibinfo
  {journal} {Phys. Rev. B}\ }\textbf {\bibinfo {volume} {95}},\ \bibinfo
  {pages} {085120} (\bibinfo {year} {2017})}\BibitemShut {NoStop}%
\bibitem [{\citenamefont {Murray}\ \emph {et~al.}(2015)\citenamefont {Murray},
  \citenamefont {Vafek},\ and\ \citenamefont {Balents}}]{PhysRevB.92.035137}%
  \BibitemOpen
  \bibfield  {author} {\bibinfo {author} {\bibfnamefont {James~M.}\
  \bibnamefont {Murray}}, \bibinfo {author} {\bibfnamefont {Oskar}\
  \bibnamefont {Vafek}}, \ and\ \bibinfo {author} {\bibfnamefont {Leon}\
  \bibnamefont {Balents}},\ }\bibfield  {title} {\enquote {\bibinfo {title}
  {Incommensurate spin density wave at a ferromagnetic quantum critical point
  in a three-dimensional parabolic semimetal},}\ }\href {\doibase
  10.1103/PhysRevB.92.035137} {\bibfield  {journal} {\bibinfo  {journal} {Phys.
  Rev. B}\ }\textbf {\bibinfo {volume} {92}},\ \bibinfo {pages} {035137}
  (\bibinfo {year} {2015})}\BibitemShut {NoStop}%
\bibitem [{\citenamefont {Boettcher}\ and\ \citenamefont
  {Herbut}(2017)}]{PhysRevB.95.075149}%
  \BibitemOpen
  \bibfield  {author} {\bibinfo {author} {\bibfnamefont {Igor}\ \bibnamefont
  {Boettcher}}\ and\ \bibinfo {author} {\bibfnamefont {Igor~F.}\ \bibnamefont
  {Herbut}},\ }\bibfield  {title} {\enquote {\bibinfo {title} {{Anisotropy
  induces non-Fermi-liquid behavior and nematic magnetic order in
  three-dimensional Luttinger semimetals}},}\ }\href {\doibase
  10.1103/PhysRevB.95.075149} {\bibfield  {journal} {\bibinfo  {journal} {Phys.
  Rev. B}\ }\textbf {\bibinfo {volume} {95}},\ \bibinfo {pages} {075149}
  (\bibinfo {year} {2017})}\BibitemShut {NoStop}%
\bibitem [{\citenamefont {Boettcher}\ and\ \citenamefont
  {Herbut}(2016)}]{PhysRevB.93.205138}%
  \BibitemOpen
  \bibfield  {author} {\bibinfo {author} {\bibfnamefont {Igor}\ \bibnamefont
  {Boettcher}}\ and\ \bibinfo {author} {\bibfnamefont {Igor~F.}\ \bibnamefont
  {Herbut}},\ }\bibfield  {title} {\enquote {\bibinfo {title} {{Superconducting
  quantum criticality in three-dimensional Luttinger semimetals}},}\ }\href
  {\doibase 10.1103/PhysRevB.93.205138} {\bibfield  {journal} {\bibinfo
  {journal} {Phys. Rev. B}\ }\textbf {\bibinfo {volume} {93}},\ \bibinfo
  {pages} {205138} (\bibinfo {year} {2016})}\BibitemShut {NoStop}%
\bibitem [{\citenamefont {Herbut}\ and\ \citenamefont
  {Janssen}(2014)}]{PhysRevLett.113.106401}%
  \BibitemOpen
  \bibfield  {author} {\bibinfo {author} {\bibfnamefont {Igor~F.}\ \bibnamefont
  {Herbut}}\ and\ \bibinfo {author} {\bibfnamefont {Lukas}\ \bibnamefont
  {Janssen}},\ }\bibfield  {title} {\enquote {\bibinfo {title} {{Topological
  Mott Insulator in Three-Dimensional Systems with Quadratic Band Touching}},}\
  }\href {\doibase 10.1103/PhysRevLett.113.106401} {\bibfield  {journal}
  {\bibinfo  {journal} {Phys. Rev. Lett.}\ }\textbf {\bibinfo {volume} {113}},\
  \bibinfo {pages} {106401} (\bibinfo {year} {2014})}\BibitemShut {NoStop}%
\bibitem [{\citenamefont {Janssen}\ and\ \citenamefont
  {Herbut}(2016)}]{PhysRevB.93.165109}%
  \BibitemOpen
  \bibfield  {author} {\bibinfo {author} {\bibfnamefont {Lukas}\ \bibnamefont
  {Janssen}}\ and\ \bibinfo {author} {\bibfnamefont {Igor~F.}\ \bibnamefont
  {Herbut}},\ }\bibfield  {title} {\enquote {\bibinfo {title} {{Excitonic
  instability of three-dimensional gapless semiconductors: Large-$N$
  theory}},}\ }\href {\doibase 10.1103/PhysRevB.93.165109} {\bibfield
  {journal} {\bibinfo  {journal} {Phys. Rev. B}\ }\textbf {\bibinfo {volume}
  {93}},\ \bibinfo {pages} {165109} (\bibinfo {year} {2016})}\BibitemShut
  {NoStop}%
\bibitem [{\citenamefont {Janssen}\ and\ \citenamefont
  {Herbut}(2017)}]{PhysRevB.95.075101}%
  \BibitemOpen
  \bibfield  {author} {\bibinfo {author} {\bibfnamefont {Lukas}\ \bibnamefont
  {Janssen}}\ and\ \bibinfo {author} {\bibfnamefont {Igor~F.}\ \bibnamefont
  {Herbut}},\ }\bibfield  {title} {\enquote {\bibinfo {title} {{Phase diagram
  of electronic systems with quadratic Fermi nodes in $2<d<4$:
  $2+\ensuremath{\epsilon}$ expansion, $4\ensuremath{-}\ensuremath{\epsilon}$
  expansion, and functional renormalization group}},}\ }\href {\doibase
  10.1103/PhysRevB.95.075101} {\bibfield  {journal} {\bibinfo  {journal} {Phys.
  Rev. B}\ }\textbf {\bibinfo {volume} {95}},\ \bibinfo {pages} {075101}
  (\bibinfo {year} {2017})}\BibitemShut {NoStop}%
\bibitem [{\citenamefont {Curnoe}(2008)}]{PhysRevB.78.094418}%
  \BibitemOpen
  \bibfield  {author} {\bibinfo {author} {\bibfnamefont {S.~H.}\ \bibnamefont
  {Curnoe}},\ }\bibfield  {title} {\enquote {\bibinfo {title} {{Structural
  distortion and the spin liquid state in
  ${\text{Tb}}_{2}{\text{Ti}}_{2}{\text{O}}_{7}$}},}\ }\href {\doibase
  10.1103/PhysRevB.78.094418} {\bibfield  {journal} {\bibinfo  {journal} {Phys.
  Rev. B}\ }\textbf {\bibinfo {volume} {78}},\ \bibinfo {pages} {094418}
  (\bibinfo {year} {2008})}\BibitemShut {NoStop}%
\bibitem [{\citenamefont {Onoda}\ and\ \citenamefont {Tanaka}(2010)}]{Onoda}%
  \BibitemOpen
  \bibfield  {author} {\bibinfo {author} {\bibfnamefont {Shigeki}\ \bibnamefont
  {Onoda}}\ and\ \bibinfo {author} {\bibfnamefont {Yoichi}\ \bibnamefont
  {Tanaka}},\ }\bibfield  {title} {\enquote {\bibinfo {title} {{Quantum Melting
  of Spin Ice: Emergent Cooperative Quadrupole and Chirality}},}\ }\href
  {\doibase 10.1103/PhysRevLett.105.047201} {\bibfield  {journal} {\bibinfo
  {journal} {Phys. Rev. Lett.}\ }\textbf {\bibinfo {volume} {105}},\ \bibinfo
  {pages} {047201} (\bibinfo {year} {2010})}\BibitemShut {NoStop}%
\bibitem [{\citenamefont {Lee}\ \emph {et~al.}(2012)\citenamefont {Lee},
  \citenamefont {Onoda},\ and\ \citenamefont {Balents}}]{SungbinBalents}%
  \BibitemOpen
  \bibfield  {author} {\bibinfo {author} {\bibfnamefont {SungBin}\ \bibnamefont
  {Lee}}, \bibinfo {author} {\bibfnamefont {Shigeki}\ \bibnamefont {Onoda}}, \
  and\ \bibinfo {author} {\bibfnamefont {Leon}\ \bibnamefont {Balents}},\
  }\bibfield  {title} {\enquote {\bibinfo {title} {Generic quantum spin ice},}\
  }\href {\doibase 10.1103/PhysRevB.86.104412} {\bibfield  {journal} {\bibinfo
  {journal} {Phys. Rev. B}\ }\textbf {\bibinfo {volume} {86}},\ \bibinfo
  {pages} {104412} (\bibinfo {year} {2012})}\BibitemShut {NoStop}%
\bibitem [{\citenamefont {Mong}\ \emph {et~al.}(2010)\citenamefont {Mong},
  \citenamefont {Essin},\ and\ \citenamefont {Moore}}]{PhysRevB.81.245209}%
  \BibitemOpen
  \bibfield  {author} {\bibinfo {author} {\bibfnamefont {Roger S.~K.}\
  \bibnamefont {Mong}}, \bibinfo {author} {\bibfnamefont {Andrew~M.}\
  \bibnamefont {Essin}}, \ and\ \bibinfo {author} {\bibfnamefont {Joel~E.}\
  \bibnamefont {Moore}},\ }\bibfield  {title} {\enquote {\bibinfo {title}
  {Antiferromagnetic topological insulators},}\ }\href {\doibase
  10.1103/PhysRevB.81.245209} {\bibfield  {journal} {\bibinfo  {journal} {Phys.
  Rev. B}\ }\textbf {\bibinfo {volume} {81}},\ \bibinfo {pages} {245209}
  (\bibinfo {year} {2010})}\BibitemShut {NoStop}%
\bibitem [{\citenamefont {Xu}\ \emph {et~al.}(2011)\citenamefont {Xu},
  \citenamefont {Weng}, \citenamefont {Wang}, \citenamefont {Dai},\ and\
  \citenamefont {Fang}}]{dWSM}%
  \BibitemOpen
  \bibfield  {author} {\bibinfo {author} {\bibfnamefont {Gang}\ \bibnamefont
  {Xu}}, \bibinfo {author} {\bibfnamefont {Hongming}\ \bibnamefont {Weng}},
  \bibinfo {author} {\bibfnamefont {Zhijun}\ \bibnamefont {Wang}}, \bibinfo
  {author} {\bibfnamefont {Xi}~\bibnamefont {Dai}}, \ and\ \bibinfo {author}
  {\bibfnamefont {Zhong}\ \bibnamefont {Fang}},\ }\bibfield  {title} {\enquote
  {\bibinfo {title} {{Chern Semimetal and the Quantized Anomalous Hall Effect
  in ${\mathrm{HgCr}}_{2}{\mathrm{Se}}_{4}$}},}\ }\href {\doibase
  10.1103/PhysRevLett.107.186806} {\bibfield  {journal} {\bibinfo  {journal}
  {Phys. Rev. Lett.}\ }\textbf {\bibinfo {volume} {107}},\ \bibinfo {pages}
  {186806} (\bibinfo {year} {2011})}\BibitemShut {NoStop}%
\bibitem [{\citenamefont {Huang}\ \emph {et~al.}(2014)\citenamefont {Huang},
  \citenamefont {Chen},\ and\ \citenamefont
  {Hermele}}]{PhysRevLett.112.167203}%
  \BibitemOpen
  \bibfield  {author} {\bibinfo {author} {\bibfnamefont {Yi-Ping}\ \bibnamefont
  {Huang}}, \bibinfo {author} {\bibfnamefont {Gang}\ \bibnamefont {Chen}}, \
  and\ \bibinfo {author} {\bibfnamefont {Michael}\ \bibnamefont {Hermele}},\
  }\bibfield  {title} {\enquote {\bibinfo {title} {{Quantum Spin Ices and
  Topological Phases from Dipolar-Octupolar Doublets on the Pyrochlore
  Lattice}},}\ }\href {\doibase 10.1103/PhysRevLett.112.167203} {\bibfield
  {journal} {\bibinfo  {journal} {Phys. Rev. Lett.}\ }\textbf {\bibinfo
  {volume} {112}},\ \bibinfo {pages} {167203} (\bibinfo {year}
  {2014})}\BibitemShut {NoStop}%
\bibitem [{\citenamefont {Li}\ \emph {et~al.}(2016)\citenamefont {Li},
  \citenamefont {Wang},\ and\ \citenamefont {Chen}}]{PhysRevB.94.201114}%
  \BibitemOpen
  \bibfield  {author} {\bibinfo {author} {\bibfnamefont {Yao-Dong}\
  \bibnamefont {Li}}, \bibinfo {author} {\bibfnamefont {Xiaoqun}\ \bibnamefont
  {Wang}}, \ and\ \bibinfo {author} {\bibfnamefont {Gang}\ \bibnamefont
  {Chen}},\ }\bibfield  {title} {\enquote {\bibinfo {title} {{Hidden multipolar
  orders of dipole-octupole doublets on a triangular lattice}},}\ }\href
  {\doibase 10.1103/PhysRevB.94.201114} {\bibfield  {journal} {\bibinfo
  {journal} {Phys. Rev. B}\ }\textbf {\bibinfo {volume} {94}},\ \bibinfo
  {pages} {201114} (\bibinfo {year} {2016})}\BibitemShut {NoStop}%
\end{thebibliography}%

\end{document}